\documentclass{aa}
\usepackage{graphicx}
\usepackage{natbib}
\usepackage[varg]{txfonts}
\usepackage{color}
\usepackage{longtable}

\usepackage{url}

\begin{document}

\title{Normal A0--A1 stars with low rotational velocities 
\thanks{Based on observations made at Observatoire de Haute-Provence (CNRS), France}\fnmsep
\thanks{Tables\,\ref{tabstars}, \ref{indivspectra}, \ref{abundances} and \ref{table_classif} are available online only, as well as the appendices}
}
\subtitle{I. Abundance determination and classification}
\author{F. Royer\inst{1} \and
M. Gebran\inst{2}    \and
R. Monier\inst{3,4}  \and
S. Adelman\inst{5}   \and
B. Smalley\inst{6}   \and\\
O. Pintado\inst{7}   \and
A. Reiners\inst{8}   \and
G. Hill\inst{9,10}   \and
A. Gulliver\inst{10}  
}
 
\institute{
GEPI/CNRS UMR 8111, Observatoire de Paris -- Universit\'e Paris Denis Diderot, 5 place Jules Janssen, 92190 Meudon, France \email{frederic.royer@obspm.fr}
\and
Department of Physics and Astronomy, Notre Dame University-Louaize, PO Box 72, Zouk Mika\"el, Lebanon
\and
LESIA/CNRS UMR 8109, Observatoire de Paris -- Universit\'e Pierre et Marie Curie -- Universit\'e Paris Denis Diderot, 5 place Jules Janssen, 92190 Meudon, France
\and
Laboratoire Lagrange, Universit\'e de Nice Sophia Antipolis, Parc Valrose, 06100 Nice, France
\and
Department of Physics, The Citadel, 171 Moultrie Street, Charleston, SC 29409, USA
\and
Astrophysics Group, Keele University, Staffordshire ST5 5BG, UK 
\and
INSUGEO-CONICET, Tucum\'an, Argentina
\and
Institut f\"ur Astrophysik G\"ottingen, Physik Fakult\"at, Friedrich-Hund-Platz 1, 37077 G\"ottingen, Germany 
\and
18A Stratford St, Auckland, New Zealand
\and
Department of Physics and Astronomy, Brandon University, Brandon, MB R7A 6A9, Canada
}
 
\date{Received 27 September 2013 / Accepted 23 December 2013}

\titlerunning{Normal A0--A1 stars with low rotational velocities I.}
\authorrunning{F. Royer et al.}

\abstract {The study of rotational velocity distributions for normal stars requires an accurate spectral characterization of the objects in order to avoid polluting the results with undetected binary or peculiar stars. This piece of information is a key issue in the understanding of the link between rotation and the presence of chemical peculiarities.} {A sample of 47 low \ensuremath{v\sin i}\ A0--A1 stars ($\ensuremath{v\sin i}<65$\ensuremath{\mbox{km}\,\mbox{s}^{-1}}), initially selected as main-sequence normal stars, are investigated with high-resolution and high signal-to-noise spectroscopic data. The aim is to detect spectroscopic binaries and chemically peculiar stars, and eventually establish a list of confirmed normal stars.} {A detailed abundance analysis and spectral synthesis is performed to derive abundances for 14 chemical species. A hierarchical classification, taking measurement errors into account, is applied to the abundance space and splits the sample into two different groups, identified as the chemically peculiar stars and the normal stars.} {We show that about one third of the sample is actually composed of spectroscopic binaries (12 double-lined and five single-lined spectroscopic binaries). The hierarchical classification breaks down the remaining sample into 13 chemically peculiar stars (or uncertain) and 17 normal stars.} {}
 \keywords{stars: early-type -- stars: rotation -- stars: abundances -- stars: chemically peculiar -- binaries: spectroscopic}
 \maketitle 


\section{Introduction}

Observations strongly suggested in the 1960s and 1970s that slow rotation is a \emph{necessary} condition for the presence of peculiarities in the spectra of A-type stars \citep[and references therein]{Prn74}.
Since then, the equatorial velocity below which chemically peculiar (hereafter CP) stars are observed is found to be around 120\,\ensuremath{\mbox{km}\,\mbox{s}^{-1}}\ \citep{AbtMod73,AbtMol95}. This observation is supported by theory that links the chemical peculiarities to the diffusion mechanism, occurring when the \ion{He}{ii} convection zone disappears at equatorial velocities lower than 70--120\,\ensuremath{\mbox{km}\,\mbox{s}^{-1}}\ \citep{Mid82}. Atomic diffusion, under the competitive action of gravitational settling and radiative accelerations, alters the chemical abundances in stellar atmospheres, when mixing motions are weak. In theory, slow rotation should be a sufficient condition for the CP phenomenon to appear and \citet{Mid80} wondered why a slowly rotating star would be non-peculiar.

The observational evidence that slow rotation is a \emph{sufficient} condition for the presence of chemical peculiarities is not as straightforward. Whereas the dichotomy between normal and CP stars according to rotation rate is rather clear and unanimous in the literature for mid to late A-type stars \citep{AbtMod73,AbtMol95,Ror_07}, this is not the case for early A-type stars, from A0 to A3.

For \citet[hereafter AM]{AbtMol95}, the bimodal shape of the equatorial velocity distribution can be explained by the different rotation rate of normal stars (fast) compared to peculiar and binary stars (slow). They conclude that the observed overlap between both distribution modes for A0--A1 stars is due to the inability to detect marginal CP stars or to evolutionary effects: rotation alone could thus explain the normal or peculiar appearance of an A star's spectrum \citep{Abt00,Adn04}.
However, authors focusing on the rotational velocities of normal stars confirm the excess of slow rotators for normal A0--A1 stars \citep{Dwy74,Raa_89,Ror_07}, when removing known CP and binary stars. This excess of slow rotators is moreover observed at higher masses \citep{ZocRor12}.
 
The nature of these objects remains unclear. There is no doubt that a fraction of them is composed of so far unidentified spectroscopic binaries and/or CP stars; just how many genuine normal stars are slow rotators seems very much open again. Are slowly rotating normal A stars young objects that will become Ap stars and do not show chemical peculiarity yet, as suggested by \citet{Abt09}? Since the spectroscopic measurement of rotational velocity (\ensuremath{v\sin i}) is a projection on the line-of-sight, are low \ensuremath{v\sin i}\ normal abundance A stars likely to be fast rotators seen at low inclination angle, such as Vega \citep{Gur_94,Hil_10}? To answer these questions, the low \ensuremath{v\sin i}\ A0--A1 normal stars from \citet[hereafter RZG]{Ror_07} need to be investigated with new high-resolution and high signal-to-noise spectroscopic data. 
The purpose of this article is to perform a detailed abundance analysis and spectral synthesis to detect potential binary or CP stars and provide a list of confirmed normal low \ensuremath{v\sin i}\ A0--A1 stars. The resulting subsample will be analyzed in greater depth in a following article. 

This paper is organized as follows: the sample, its selection and the spectroscopic observations are described in Sect.\,\ref{sect_sample}. 
Section\,\ref{sect_rv} deals with the determination of radial velocities and gives a list of suspected binaries. The atmospheric parameters and rotational velocities are respectively derived in Sect.\,\ref{sect_params} and Sect.\,\ref{sect_vsini}. Section\,\ref{sect_abund} presents the determination of abundance patterns and Sect.\,\ref{sect_classif} details the classification based on these chemical abundances. The abundance patterns and the rotational velocity distribution are discussed in Sect.\,\ref{sect_discussion} and the results are summarized in Sect.\,\ref{sect_conclusion}. Comments on individual stars are given in Appendix\,\ref{comments}.

\section{Data sample}
\label{sect_sample}
\subsection{Target selection}
RZG built a sample of main-sequence A-type stars with homogenized \ensuremath{v\sin i}\ using values from AM and \citet{Ror_02a,Ror_02b}. Chemically peculiar stars were removed on the basis of the spectral classification and the catalog of \citet{Ren_91}. Binary stars were discarded on the basis of HIPPARCOS data \citep{Hip} and the catalog of spectroscopic binaries from \citet{Pet_85}. For A0--A1 stars, these criteria reduced the size of the original sample by about one third.

 This paper focuses on a subsample of the A0--A1 normal stars, selected on their low \ensuremath{v\sin i}\ ($\le 65$\,\ensuremath{\mbox{km}\,\mbox{s}^{-1}}), in order to investigate the slow rotator part of the distribution and accurately check whether these stars are normal or harbor signatures of multiplicity and/or chemical peculiarity. Using this criterion, 73 stars are selected on the whole sky. Among them, 47 can be observed from Observatoire de Haute-Provence (OHP). 
 Table\,\ref{tabstars}, available online, lists the 47 targets defining our sample, together with their spectral type, $V$ magnitude, \ensuremath{v\sin i}\ \citep{Ror_02b} and the different parameters derived in the next sections.
\onllongtab{
\begin{table*}[!ht]
\begin{center}
\caption{List of the 47 targets, with their derived parameters: rotational velocities, microturbulent velocity $\xi_\mathrm{t}$, effective temperature, gravity, luminosity and limb-darkening coefficient $\epsilon$ in the $B$ band.}
\setlength{\tabcolsep}{5pt}
\begin{tabular}{rllrrrcrccc}
\hline\hline
   HD~& SpType   &\multicolumn{1}{c}{$V$~~~~}&  \multicolumn{3}{c}{\ensuremath{v\sin i}} & $\xi_\mathrm{t}$& \ensuremath{T_\mathrm{eff}} & $\log g$ & $\log L/L_\odot$ & $\epsilon$ \\		 
       &          & (mag)        & \multicolumn{3}{c}{(\ensuremath{\mbox{km}\,\mbox{s}^{-1}})} &  (\ensuremath{\mbox{km}\,\mbox{s}^{-1}})         & (K)   &        &  \\ 				    
\cline{4-6}
&              &                     &  \textsc{rgbgz}   & \multicolumn{1}{c}{\textsc{synth}}  &  \multicolumn{1}{c}{\textsc{ft}}       &		       &       &	  	\\				    
\hline	    
  1439 & A0IV	& 5.875    &  39 &  43.0  & $ 45.3{\scriptstyle\pm1.5}$ & 1.0  &  9640  &  3.75   &  $2.08{\scriptstyle\pm0.04}$&  0.6145 \\  
  1561 & A0Vs	& 6.538    &  60 &  63.7  & $ 66.2{\scriptstyle\pm1.4}$ & 1.3  &  8860  &  3.35   &  $1.91{\scriptstyle\pm0.07}$&  0.6580 \\  
  6530 & A1V	& 5.578    &  51 &  48.8  & $103.1{\scriptstyle\pm3.0}$ & 1.6  &  9500  &  3.87   &  $2.25{\scriptstyle\pm0.04}$&  0.6191 \\  
 20149 & A1Vs	& 5.606    &  23 &  21.8  & $ 21.9{\scriptstyle\pm1.0}$ & 1.0  &  9640  &  4.04   &  $2.15{\scriptstyle\pm0.06}$&  0.6113 \\  
 21050 & A1V	& 6.070    &  27 &  26.5  & $ 27.8{\scriptstyle\pm0.9}$ & 1.4  & 10420  &  4.30   &  $1.62{\scriptstyle\pm0.03}$&  0.5781 \\  
 25175 & A0V	& 6.311    &  55 &  55.0  & $ 56.7{\scriptstyle\pm1.5}$ & 1.5  &  9920  &  3.75   &  $2.09{\scriptstyle\pm0.07}$&  0.6031 \\  
 28780 & A1V	& 5.908    &  28 &  31.0  & $ 33.0{\scriptstyle\pm0.7}$ & 1.7  &  9640  &  3.86   &  $1.88{\scriptstyle\pm0.05}$&  0.6135 \\   
 30085 & A0IV	& 6.345    &  26 &  23.0  & $ 24.2{\scriptstyle\pm0.6}$ & 0.5  & 11300  &  3.95   &  $2.28{\scriptstyle\pm0.07}$&  0.5516 \\   
 33654 & A0V	& 6.156    &  60 &  72.0  & $ 73.0{\scriptstyle\pm4.1}$ & 3.3  &  9220  &  2.90   &  $3.63{\scriptstyle\pm0.84}$&  0.6389 \\   
 39985 & A0IV	& 5.971    &  28 &  28.5  & $ 28.2{\scriptstyle\pm1.2}$ & 0.6  & 10240  &  3.62   &  $1.83{\scriptstyle\pm0.05}$&  0.5917 \\   
 40446 & A1Vs	& 5.213    &  27 &  28.0  & $ 23.6{\scriptstyle\pm3.5}$ & 0.5  &  9550	&  3.95	  &	 $2.16{\scriptstyle\pm0.10}$&  0.6161 \\   
 46642 & A0Vs	& 6.46     &  49 &  57.0  & $ 57.7{\scriptstyle\pm1.4}$ & 2.3  &  9890  &  3.89   &  $1.57{\scriptstyle\pm0.11}$&  0.6028 \\   
 47863 & A1V	& 6.284    &  40 &  41.0  & $ 47.1{\scriptstyle\pm2.6}$ & 2.2  &  9520  &  3.45   &  $2.34{\scriptstyle\pm0.10}$&  0.6224 \\ 
 50931 & A0V	& 6.274    &  65 &  75.0  & $ 84.0{\scriptstyle\pm5.5}$ & 1.4  &  9440  &  4.13   &  $1.31{\scriptstyle\pm0.03}$&  0.6185 \\ 
 58142 & A1V	& 4.614    &  19 &  19.0  & $ 18.7{\scriptstyle\pm0.4}$ & 1.7  &  9520  &  3.79   &  $1.97{\scriptstyle\pm0.02}$&  0.6193 \\   
 65900 & A1V	& 5.646    &  33 &  34.8  & $ 36.4{\scriptstyle\pm1.5}$ & 2.0  &  9600  &  4.01   &  $1.67{\scriptstyle\pm0.04}$&  0.6135 \\   
 67959 & A1V	& 6.217    &  18 &  16.7  & $ 15.5{\scriptstyle\pm0.5}$ & 2.0  &  9310  &  3.71   &  $2.03{\scriptstyle\pm0.07}$&  0.6296 \\   
 72660 & A1V	& 5.799    &   9 &   7.0  & $  6.5{\scriptstyle\pm0.4}$ & 1.5  &  9640  &  4.03   &  $1.57{\scriptstyle\pm0.03}$&  0.6116 \\   
 73316 & A1V	& 6.542    &  33 &  32.8  & $ 35.2{\scriptstyle\pm1.1}$ & 2.0  &  9830  &  4.30   &  $1.49{\scriptstyle\pm0.04}$&  0.5994 \\  
 83373 & A1V	& 6.391    &  28 &  28.5  & $ 29.8{\scriptstyle\pm1.4}$ & 1.6  & 10200  &  4.10   &  $1.73{\scriptstyle\pm0.05}$&  0.5884 \\
 85504 & A0Vs	& 6.02     &  27 &  25.7  & $ 27.6{\scriptstyle\pm2.2}$ & 1.6  & 10200  &  3.82   &  $2.26{\scriptstyle\pm0.08}$&  0.5915 \\   
 89774 & A1V	& 6.169    &  60 &  63.3  & $ 65.5{\scriptstyle\pm2.1}$ & 2.1  &  9630  &  3.90   &  $1.83{\scriptstyle\pm0.04}$&  0.6135 \\ 
 95418 & A1V	& 2.346    &  46 &  45.7  & $ 46.2{\scriptstyle\pm1.2}$ & 2.2  &  9620  &  3.89   &  $1.82{\scriptstyle\pm0.00}$&  0.6140 \\ 
101369 & A0V	& 6.210    &  65 &  67.5  & $ 70.0{\scriptstyle\pm3.9}$ & 1.0  &  9700  &  3.82   &  $1.97{\scriptstyle\pm0.11}$&  0.6112 \\ 
104181 & A1V	& 5.357    &  44 &  59.0  & $ 55.5{\scriptstyle\pm1.6}$ & 1.0  &  9660  &  4.00	  &  $1.79{\scriptstyle\pm0.02}$&  0.6110 \\ 
107655 & A0V	& 6.176    &  46 &  45.0  & $ 47.2{\scriptstyle\pm1.4}$ & 2.0  &  9680  &  4.10   &  $1.53{\scriptstyle\pm0.04}$&  0.6088 \\ 
119537 & A1V	& 6.502    &  13 &  16.7  & $ 13.3{\scriptstyle\pm0.7}$ & 2.3  &  9260  &  4.14   &  $1.54{\scriptstyle\pm0.04}$&  0.6269 \\ 
127304 & A0Vs	& 6.055    &  14 &   8.9  & $  7.7{\scriptstyle\pm0.6}$ & 1.5  & 10050  &  4.11   &  $1.70{\scriptstyle\pm0.04}$&  0.5939 \\ 
132145 & A1V	& 6.506    &  15 &  13.5  & $ 12.7{\scriptstyle\pm0.6}$ & 1.7  &  9680  &  4.25   &  $1.59{\scriptstyle\pm0.05}$&  0.6063 \\ 
133962 & A1V	& 5.581    &  49 &  52.3  & $ 54.8{\scriptstyle\pm1.6}$ & 1.2  & 10130  &  4.32   &  $1.66{\scriptstyle\pm0.02}$&  0.5882 \\ 
145647 & A0V	& 6.092    &  43 &  44.0  & $ 47.2{\scriptstyle\pm1.3}$ & 0.5  &  9560  &  3.95   &  $1.72{\scriptstyle\pm0.05}$&  0.6159 \\ 
145788 & A1V	& 6.255    &  16 &  13.0  & $  9.8{\scriptstyle\pm0.8}$ & 2.3  &  9410  &  3.73   &  $1.88{\scriptstyle\pm0.08}$&  0.6250 \\ 
154228 & A1V	& 5.918    &  42 &  44.8  & $ 45.2{\scriptstyle\pm1.2}$ & 2.2  &  9750  &  4.20   &  $1.40{\scriptstyle\pm0.02}$&  0.6042 \\ 
156653 & A1V	& 6.002    &  43 &  43.6  & $ 45.3{\scriptstyle\pm2.0}$ & 2.3  &  9270  &  3.69   &  $1.82{\scriptstyle\pm0.08}$&  0.6320 \\ 
158716 & A1V	& 6.464    &  15 &   8.0  & $  6.4{\scriptstyle\pm0.8}$ & 2.8  &  9300  &  4.39   &  $1.19{\scriptstyle\pm0.03}$&  0.6215 \\ 
172167 & A0V	& 0.03     &  24 &  23.5  & $ 24.5{\scriptstyle\pm1.4}$ & 1.0  &  9550  &  4.05   &  $1.73{\scriptstyle\pm0.00}$&  0.6149 \\ 
174567 & A0Vs	& 6.634    &  15 &  13.0  & $ 10.5{\scriptstyle\pm0.9}$ & 1.0  &  9710  &  3.59   &  $2.23{\scriptstyle\pm0.10}$&  0.6131 \\ 
176984 & A1V	& 5.42     &  23 &  30.0  & $ 29.6{\scriptstyle\pm1.0}$ & 2.0  &  9680  &  3.44   &  $2.20{\scriptstyle\pm0.04}$&  0.6155 \\ 
183534 & A1V	& 5.75     &  49 &  53.0  & $ 51.8{\scriptstyle\pm5.8}$ & 0.2  &  9930  &  4.24   &  $1.79{\scriptstyle\pm0.02}$&  0.5966 \\ 
196724 & A0V	& 4.82     &  52 &  50.5  & $ 51.6{\scriptstyle\pm2.2}$ & 1.3  & 10400  &  4.18   &  $1.87{\scriptstyle\pm0.06}$&  0.5801 \\ 
198552 & A1Vs	& 6.618    &  52 &  51.6  & $ 53.7{\scriptstyle\pm1.1}$ & 3.0  &  9230  &  4.14   &  $1.39{\scriptstyle\pm0.06}$&  0.6284 \\ 
199095 & A0V	& 5.748    &  32 &  27.8  & $ 27.8{\scriptstyle\pm1.3}$ & 0.9  &  9920  &  4.05   &  $1.79{\scriptstyle\pm0.02}$&  0.5999 \\ 
217186 & A1V	& 6.342    &  60 &  64.0  & $ 66.7{\scriptstyle\pm2.0}$ & 1.9  &  9190  &  4.00	  &  $1.29{\scriptstyle\pm0.04}$&  0.6330 \\ 
219290 & A0V	& 6.319    &  54 &  59.0  & $ 61.4{\scriptstyle\pm1.9}$ & 1.5  &  9790  &  4.15   &  $1.66{\scriptstyle\pm0.04}$&  0.6034 \\ 
219485 & A0V	& 5.886    &  23 &  27.0  & $ 27.3{\scriptstyle\pm0.8}$ & 1.2  &  9580  &  3.81   &  $1.91{\scriptstyle\pm0.03}$&  0.6165 \\ 
223386 & A0V	& 6.328    &  33 &  33.0  & $ 36.8{\scriptstyle\pm1.2}$ & 1.2  &  9890  &  4.11   &  $1.62{\scriptstyle\pm0.07}$&  0.6001 \\ 
223855 & A1V	& 6.292    &  60 &  62.0  & $ 71.5{\scriptstyle\pm4.6}$ & 1.3  &  9890  &  4.14   &  $1.70{\scriptstyle\pm0.06}$&  0.5999 \\ 
\hline
\label{tabstars}
\end{tabular}
\tablefoot{The spectral types and $V$ magnitudes are taken from the HIPPARCOS catalog \citep{Hip}. 
The listed \ensuremath{v\sin i}\ values are: (RGBGZ) taken from \citet{Ror_02b}, with an error of $\pm10$\%, (SYNTH) derived from the spectral synthesis detailed in Sect.\,\ref{sect_params}, with a typical error of $\pm5$\% and (FT) derived from the Fourier analysis of individual line profiles (Sect.\,\ref{sect_vsini}). Errors on $\xi_\mathrm{t}$, \ensuremath{T_\mathrm{eff}}\ and $\log g$ are respectively $\pm0.5$\,\ensuremath{\mbox{km}\,\mbox{s}^{-1}}, $\pm125$\,K and $\pm0.2$\,dex.}
\end{center}
\end{table*}  
}

\subsection{Spectroscopic observations}
Spectra of our targets were collected at OHP and observations were spread over an eight year period. The first two runs (April 2005 and June 2006) used the \'ELODIE spectrograph ($R \approx 42000$, \citealt{Bae_96}). Then \'ELODIE was decommissioned in August 2006 and replaced with a more efficient instrument offering a higher spectral resolution: SOPHIE ($R \approx 75000$, \citealt{Pet_08}). The last part of our program used SOPHIE, in three different runs: July 2009, February 2011 and February 2012. Archival data have also been used: about half the \'ELODIE spectra were taken from the \'ELODIE archive\footnote{\url{http://atlas.obs-hp.fr/elodie}} \citep{Moa_04} and spectra of Vega were taken from the SOPHIE archive\footnote{\url{http://atlas.obs-hp.fr/sophie}}. Table\,\ref{indivspectra} (available electronically) lists the different observations of our targets, indicates the corresponding instrument, the observation date, the number of co-added spectra, the modified Julian date at the center of the exposure(s), the signal-to-noise ratio (S/N) derived using the DER\_SNR algorithm \citep{Str_08} and the measured radial velocity, corrected from the barycentric motion (see Sect.\,\ref{sect_rv}).
The initial observational strategy was derived from the twofold goal of our program: (i) obtain good S/N spectra ($\approx 150$--200) to perform, using synthetic spectra at low \ensuremath{v\sin i}, a detailed abundance analysis of unblended weak lines, allowing us to identify the normal stars; (ii) focus on these stars and obtain high S/N spectra ($\approx 400$) to analyze the line profiles and search for gravity-darkening signatures.
  
\subsection{Data reduction}

For both \'ELODIE and SOPHIE, data are automatically reduced to produce 1D extracted and wavelength calibrated \'echelle orders. When stars are observed several times during one observation night, the corresponding spectra are coadded.
Then, for each reduced spectrum, \'echelle orders are normalized separately, using a Chebychev polynomial
fit with sigma clipping, rejecting points above 6-$\sigma$ or below 1-$\sigma$ of the continuum. Normalized orders are merged together, weighted by the blaze function and resampled in a constant wavelength step $\Delta\lambda=0.02$\AA. Only the spectral intervals outside the wings of Balmer lines and the atmospheric telluric bands are finally retained: 4150--4300\AA, 4400--4790\AA, 4920--5850\AA\ and 6000--6275\AA. 

\begin{figure}[!t]
\centering
\resizebox{\hsize}{!}{\includegraphics{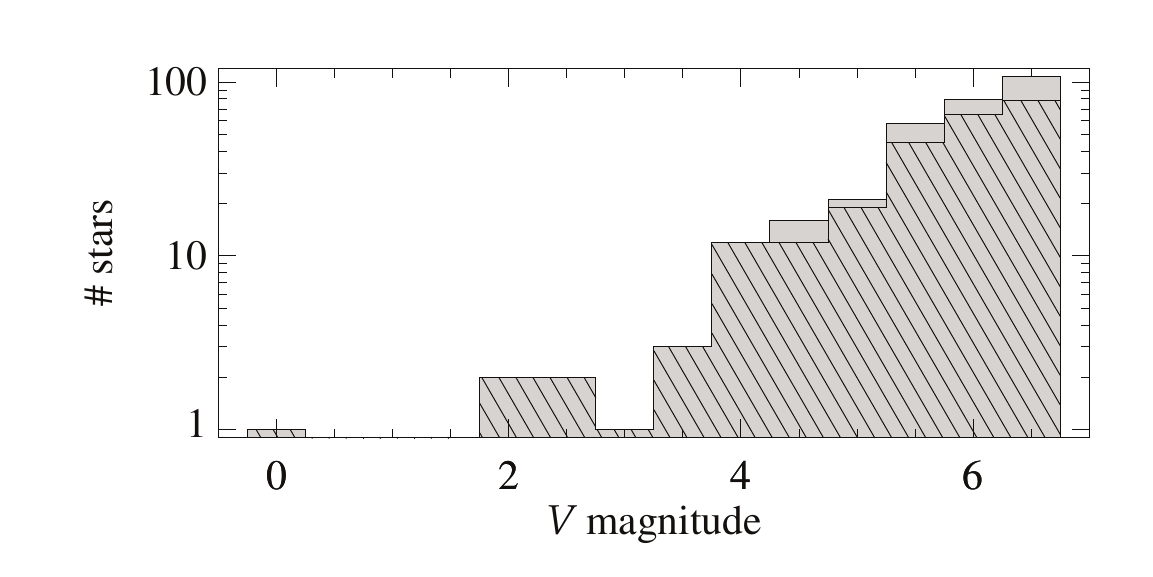}}
\caption{Counts in $V$ magnitude bins for northern ($\delta>-15\degr$) A0--A1 main sequence stars in the HIPPARCOS catalog (gray) and in the \ensuremath{v\sin i}\ sample (hatched).}
\label{complet}
\end{figure}

\subsection{Completeness}
 \label{completeness}
 Besides selections in \ensuremath{v\sin i}\ and the absence of spectroscopic peculiarities, the aforedescribed sample is magnitude-limited and censored in declination and spectral type. These censorships can be applied to the HIPPARCOS catalog \citep{Hip}, complete down to $V=7.3$\, \citep{Pen_97}, to estimate the completeness of the sample.
The selection criteria are the following:
\begin{itemize}
\item declination $\delta$ higher than $-15\degr$ to reproduce the observability bias due to the location of OHP (i.e. 63\% of the celestial sphere),
\item spectral class containing A0 or A1, and luminosity class either V, IV/V or IV, to reproduce the selection made in RZG,
\item magnitude $V$ brighter than 6.65\,.
\end{itemize}
This selection results in 303
stars from the HIPPARCOS catalog, among them 240 belong to the sample studied by RZG. The star counts per bin of $V$-magnitude are compared in Fig.\,\ref{complet}.  The ratio of these two counts gives a completeness of about 80\%. 
 
When limited to normal stars, using the results from RZG, 151 stars remain out of the 240. Our 47 targets correspond to the \ensuremath{v\sin i}-truncated subsample out of these 151 stars.

\onllongtab{
\begin{longtable}{rlcclrrl}
\caption{List of observations and barycentric radial velocity measurements (RV) for the target sample. Asterisks, following the instrument name, indicate the spectra have been retrieved from the corresponding archive. Spectra observed the same night were combined into one spectrum. The number of input spectra, the modified Julian date (MJD) at the center of the exposure(s) and the combined signal-to-noise (S/N) are also  given.\label{indivspectra}}\\
\hline\hline
HD~     & Instrument & Date &  \#	&  \multicolumn{1}{c}{MJD} & S/N & \multicolumn{1}{c}{RV}  \\ 
        &		&	&spectra&      &     & \multicolumn{1}{c}{(\ensuremath{\mbox{km}\,\mbox{s}^{-1}})}&	 \\  
\hline
\endfirsthead
\caption{continued.}\\
\hline\hline
HD~     & Instrument & Date &  \#	&  \multicolumn{1}{c}{MJD} & S/N & \multicolumn{1}{c}{RV}  \\ 
        &		&	&spectra&      &     & \multicolumn{1}{c}{(\ensuremath{\mbox{km}\,\mbox{s}^{-1}})}&	 \\  
\hline
\endhead
\hline
\endfoot

  1439 & SOPHIE   & 2009-08-05   & 1  & 55049.055625   &  213 & $  -9.57 \pm   0.46$ \\
       & SOPHIE   & 2011-02-11   & 2  & 55603.765526   &  326 & $ -10.38 \pm   0.48$ \\
\hline
  1561 & SOPHIE   & 2009-08-05   & 1  & 55049.042685   &  210 & $ -18.44 \pm   0.88$ \\
       & SOPHIE   & 2011-02-11   & 3  & 55603.792789   &  337 & $  -0.04 \pm   0.92$ \\
\hline
  6530 & SOPHIE   & 2009-08-05   & 3  & 55049.122963   &  387 & $  10.71 \pm   2.75$ \\
\hline
 20149 & ELODIE*  & 2004-01-02   & 1  & 53006.871601   &  106 & $  -9.01 \pm   0.16$ \\
       & SOPHIE   & 2009-08-05   & 3  & 55049.086474   &  399 & $ -12.11 \pm   0.16$ \\
       & SOPHIE   & 2011-02-11   & 2  & 55603.834201   &  356 & $ -12.32 \pm   0.16$ \\
       & SOPHIE   & 2012-02-13   & 2  & 55970.794641   &  341 & $ -10.70 \pm   0.16$ \\
       & SOPHIE   & 2012-02-14   & 2  & 55971.764421   &  370 & $ -10.88 \pm   0.16$ \\
\hline
 21050 & SOPHIE   & 2009-08-05   & 1  & 55049.104537   &  226 & $  -1.10 \pm   0.23$ \\
       & SOPHIE   & 2011-02-11   & 2  & 55603.856759   &   69 & $  -1.08 \pm   0.44$ \\
       & SOPHIE   & 2012-02-13   & 2  & 55970.773571   &  328 & $  -0.89 \pm   0.24$ \\
\hline
 25175 & SOPHIE   & 2011-02-11   & 1  & 55603.817766   &  222 & $  31.62 \pm   0.76$ \\
       & SOPHIE   & 2012-02-13   & 3  & 55970.818221   &  308 & $  31.33 \pm   0.73$ \\
\hline
 28780 & SOPHIE   & 2012-02-13   & 2  & 55970.847066   &  340 & $ -14.19 \pm   0.29$ \\
\hline
 30085 & SOPHIE   & 2012-02-13   & 3  & 55970.874525   &  316 & $   8.27 \pm   0.20$ \\
\hline
 33654 & SOPHIE   & 2012-02-13   & 2  & 55970.904277   &  180 & $  -3.12 \pm   1.10$ \\
\hline
 39985 & ELODIE*  & 1995-12-20   & 1  & 50071.994159   &  138 & $  13.47 \pm   0.24$ \\
\hline
 40446 & ELODIE*  & 1999-12-20   & 1  & 51532.843002   &  160 & $  47.80 \pm   0.20$ \\
\hline
 46642 & ELODIE*  & 2003-01-15   & 1  & 52654.978756   &  190 & $  31.62 \pm   0.81$ \\
       & SOPHIE   & 2012-02-14   & 3  & 55971.822326   &  332 & $  38.41 \pm   0.78$ \\
\hline
 47863 & SOPHIE   & 2012-02-14   & 3  & 55971.789437   &  365 & $  23.80 \pm   0.52$ \\
\hline
 50931 & SOPHIE   & 2012-02-14   & 3  & 55971.859687   &  345 & $  42.06 \pm   1.46$ \\
\hline
 58142 & ELODIE*  & 1998-01-28   & 1  & 50841.980507   &  172 & $  26.94 \pm   0.12$ \\
       & ELODIE*  & 2004-01-03   & 1  & 53007.964181   &  300 & $  26.76 \pm   0.12$ \\
       & ELODIE*  & 2005-02-02   & 1  & 53403.947654   &  270 & $  26.90 \pm   0.12$ \\
       & ELODIE*  & 2005-02-03   & 1  & 53404.944216   &  292 & $  26.66 \pm   0.12$ \\
\hline
 65900 & ELODIE*  & 1995-12-21   & 1  & 50072.040836   &  107 & $  43.06 \pm   0.36$ \\
\hline
 67959 & ELODIE   & 2005-04-21   & 1  & 53481.846784   &  106 & $  23.74 \pm   0.09$ \\
\hline
 72660 & ELODIE*  & 1997-03-20   & 1  & 50527.808905   &   64 & $   3.22 \pm   0.03$ \\
       & ELODIE*  & 2004-01-03   & 1  & 53007.074899   &  118 & $   4.28 \pm   0.06$ \\
       & ELODIE*  & 2004-01-04   & 1  & 53009.017029   &  152 & $   4.68 \pm   0.06$ \\
       & ELODIE*  & 2004-04-11   & 1  & 53106.829269   &  192 & $   4.74 \pm   0.07$ \\
\hline
 73316 & SOPHIE   & 2012-02-13   & 3  & 55970.946798   &  146 & $  27.68 \pm   0.34$ \\
       & SOPHIE   & 2012-02-14   & 3  & 55971.894159   &  335 & $  28.02 \pm   0.33$ \\
\hline
 83373 & SOPHIE   & 2012-02-14   & 4  & 55971.999699   &  292 & $  26.58 \pm   0.24$ \\
\hline
 85504 & ELODIE*  & 1996-04-24   & 1  & 50197.825920   &  114 & $ 103.76 \pm   0.24$ \\
       & SOPHIE   & 2012-02-13   & 2  & 55971.001111   &   63 & $ 104.03 \pm   0.32$ \\
       & SOPHIE   & 2012-02-14   & 4  & 55971.960573   &  289 & $ 103.70 \pm   0.24$ \\
\hline
 89774 & ELODIE   & 2005-04-22   & 1  & 53482.912204   &  102 & $  13.28 \pm   0.94$ \\
       & SOPHIE   & 2012-02-14   & 3  & 55971.926975   &  293 & $  12.18 \pm   0.85$ \\
\hline
 95418 & ELODIE*  & 1997-02-20   & 1  & 50499.547959   &  170 & $ -13.52 \pm   0.66$ \\
       & ELODIE*  & 2004-03-10   & 1  & 53074.002745   &  232 & $ -13.73 \pm   0.50$ \\
       & ELODIE*  & 2005-02-04   & 1  & 53405.056783   &  283 & $ -13.89 \pm   0.53$ \\
       & SOPHIE   & 2011-02-11   & 5  & 55604.191435   &  375 & $ -13.67 \pm   0.51$ \\
\hline
101369 & ELODIE   & 2005-04-21   & 2  & 53481.910739   &  182 & $   7.70 \pm   1.79$ \\
\hline
104181 & ELODIE*  & 2004-04-26   & 1  & 53121.940547   &  209 & $  -1.29 \pm   0.84$ \\
       & SOPHIE   & 2012-02-14   & 3  & 55972.076979   &  299 & $  -0.34 \pm   0.73$ \\
\hline
107655 & ELODIE*  & 2004-04-08   & 1  & 53103.058789   &  243 & $  -2.74 \pm   0.59$ \\
       & SOPHIE   & 2012-02-14   & 3  & 55972.034603   &  338 & $  -3.49 \pm   0.54$ \\
\hline
119537 & ELODIE   & 2006-06-02   & 1  & 53888.856696   &  105 & $  54.81 \pm   0.08$ \\
       & SOPHIE   & 2012-02-14   & 7  & 55972.175921   &  254 & $ -62.46 \pm   0.07$ \\
\hline
127304 & ELODIE   & 2005-04-21   & 1  & 53481.985365   &  223 & $ -22.30 \pm   0.11$ \\
\hline
132145 & ELODIE   & 2005-04-22   & 2  & 53482.066887   &  219 & $  -9.98 \pm   0.07$ \\
       & SOPHIE   & 2012-02-14   & 4  & 55972.113021   &  277 & $  -9.73 \pm   0.07$ \\
\hline
133962 & ELODIE   & 2005-04-22   & 1  & 53482.131317   &  201 & $ -12.63 \pm   0.73$ \\
       & SOPHIE   & 2011-02-11   & 2  & 55604.224086   &  215 & $ -13.29 \pm   0.73$ \\
       & SOPHIE   & 2012-02-14   & 3  & 55972.058538   &  329 & $ -13.11 \pm   0.72$ \\
\hline
145647 & ELODIE   & 2006-05-31   & 2  & 53886.908366   &  107 & $ -15.12 \pm   0.85$ \\
       & ELODIE   & 2006-06-03   & 1  & 53889.067390   &  156 & $ -15.75 \pm   0.58$ \\
\hline
145788 & ELODIE   & 2006-06-01   & 2  & 53887.968344   &  128 & $ -13.05 \pm   0.05$ \\
\hline
154228 & ELODIE*  & 1999-06-05   & 1  & 51334.988236   &  163 & $ -31.10 \pm   0.50$ \\
       & ELODIE*  & 1999-06-06   & 1  & 51335.970245   &  244 & $ -31.66 \pm   0.47$ \\
\hline
156653 & ELODIE*  & 1996-04-26   & 1  & 50199.123847   &  113 & $   2.37 \pm   0.47$ \\
       & SOPHIE   & 2009-08-05   & 3  & 55048.835204   &  294 & $   4.24 \pm   0.46$ \\
\hline
158716 & ELODIE   & 2006-06-02   & 2  & 53888.924720   &  117 & $ -25.44 \pm   0.03$ \\
\hline
172167 & ELODIE*  & 2004-02-28   & 1  & 53063.215076   &  266 & $ -13.91 \pm   0.19$ \\
       & ELODIE*  & 2004-03-01   & 1  & 53065.215871   &  288 & $ -13.82 \pm   0.20$ \\
       & ELODIE*  & 2004-03-27   & 1  & 53091.176158   &  275 & $ -13.99 \pm   0.20$ \\
       & ELODIE*  & 2004-03-29   & 2  & 53093.148334   &  357 & $ -13.91 \pm   0.19$ \\
       & ELODIE*  & 2004-08-30   & 2  & 53247.816155   &  378 & $ -13.68 \pm   0.19$ \\
       & ELODIE*  & 2004-08-31   & 2  & 53248.796281   &  355 & $ -13.62 \pm   0.19$ \\
       & ELODIE*  & 2004-09-23   & 2  & 53271.772018   &  317 & $ -13.77 \pm   0.19$ \\
       & ELODIE*  & 2004-09-24   & 2  & 53272.771336   &  320 & $ -13.50 \pm   0.19$ \\
       & ELODIE*  & 2004-09-25   & 2  & 53273.769548   &  347 & $ -13.60 \pm   0.20$ \\
       & ELODIE*  & 2004-11-05   & 2  & 53314.738914   &  296 & $ -13.52 \pm   0.19$ \\
       & ELODIE*  & 2004-11-06   & 1  & 53315.726548   &  246 & $ -13.86 \pm   0.18$ \\
       & ELODIE*  & 2004-11-07   & 2  & 53316.740124   &  251 & $ -13.49 \pm   0.20$ \\
       & ELODIE*  & 2004-11-09   & 1  & 53318.732878   &  152 & $ -14.01 \pm   0.20$ \\
       & ELODIE*  & 2005-04-20   & 2  & 53480.107426   &  309 & $ -13.89 \pm   0.19$ \\
       & ELODIE*  & 2005-05-24   & 2  & 53514.082351   &  301 & $ -13.79 \pm   0.19$ \\
       & ELODIE*  & 2005-05-25   & 1  & 53515.036852   &  254 & $ -13.94 \pm   0.20$ \\
       & SOPHIE*  & 2006-09-08   & 1  & 53987.400894   &  109 & $ -14.07 \pm   0.21$ \\
       & SOPHIE*  & 2009-09-28   & 4  & 55102.762360   & 1055 & $ -13.81 \pm   0.19$ \\
       & SOPHIE*  & 2009-09-29   & 1  & 55103.761169   &  467 & $ -13.81 \pm   0.19$ \\
       & SOPHIE*  & 2009-09-30   & 1  & 55104.763217   &  422 & $ -13.72 \pm   0.19$ \\
       & SOPHIE*  & 2009-10-06   & 1  & 55110.777199   &  386 & $ -13.69 \pm   0.19$ \\
       & SOPHIE*  & 2009-10-07   & 2  & 55111.751158   &  443 & $ -13.76 \pm   0.19$ \\
       & SOPHIE*  & 2009-10-08   & 1  & 55112.772696   &  438 & $ -13.68 \pm   0.19$ \\
       & SOPHIE*  & 2009-10-09   & 1  & 55113.756793   &  405 & $ -13.66 \pm   0.19$ \\
       & SOPHIE*  & 2009-10-10   & 3  & 55114.744915   &  571 & $ -13.71 \pm   0.19$ \\
       & SOPHIE*  & 2009-10-11   & 1  & 55115.742314   &  300 & $ -13.83 \pm   0.19$ \\
\hline
174567 & ELODIE   & 2006-05-31   & 2  & 53887.000591   &  119 & $ -14.18 \pm   0.05$ \\
       & SOPHIE   & 2009-08-05   & 3  & 55048.871358   &  355 & $ -10.84 \pm   0.05$ \\
\hline
176984 & ELODIE   & 2006-06-02   & 1  & 53888.998090   &  153 & $ -47.68 \pm   0.29$ \\
       & SOPHIE   & 2009-08-05   & 3  & 55048.903823   &  402 & $ -43.75 \pm   0.24$ \\
\hline
183534 & ELODIE   & 2006-06-01   & 1  & 53887.075317   &  148 & $   9.88 \pm   0.63$ \\
\hline
196724 & ELODIE*  & 2004-11-12   & 1  & 53321.779634   &  301 & $ -21.97 \pm   0.64$ \\
       & ELODIE   & 2006-06-02   & 1  & 53888.083812   &  156 & $ -18.31 \pm   0.52$ \\
\hline
198552 & SOPHIE   & 2009-08-05   & 2  & 55048.952500   &  314 & $   5.47 \pm   0.69$ \\
\hline
199095 & ELODIE*  & 2005-10-14   & 1  & 53657.768627   &  168 & $ -24.99 \pm   0.21$ \\
       & SOPHIE   & 2009-08-05   & 3  & 55048.925880   &  325 & $   3.49 \pm   0.23$ \\
\hline
217186 & SOPHIE   & 2009-08-05   & 3  & 55048.980456   &  287 & $  18.67 \pm   1.34$ \\
\hline
219290 & SOPHIE   & 2009-08-05   & 2  & 55049.003953   &  253 & $ -16.88 \pm   0.83$ \\
\hline
219485 & SOPHIE   & 2009-08-05   & 1  & 55049.016284   &  239 & $  -4.48 \pm   0.21$ \\
\hline
223386 & SOPHIE   & 2009-08-05   & 1  & 55049.027685   &  235 & $ -18.57 \pm   0.34$ \\
\hline
223855 & SOPHIE   & 2009-08-05   & 1  & 55049.067083   &  225 & $   3.68 \pm   1.05$ \\

\hline
\end{longtable}
}
   
  \section{Radial velocities}
\label{sect_rv}

The normalized spectra are cross-correlated with a synthetic template extracted from the POLLUX database\footnote{\url{http://pollux.graal.univ-montp2.fr}} \citep{Pas_10} corresponding to the parameters $\ensuremath{T_\mathrm{eff}}=9500$\,K, $\ensuremath{\log g}=4$ and solar metallicity \citep[computed with \textsc{synspec48},][]{HuyLaz92}, to compute the cross-correlation function (hereafter CCF). The radial velocity is derived from the 
parabolic fit of the upper part (10\%) of the CCF and the values are given in Table\,\ref{indivspectra}.
The error on the radial velocity is determined from the cross-correlation function using the formulation given by \citet{Zur03} and assuming the parabolic fit of the CCF.

\subsection{Combining \'ELODIE and SOPHIE}

Fourteen of our targets have spectra collected with both \'ELODIE and SOPHIE. In order to combine the different velocities and detect possible variations, the radial velocity offset between both instruments has to be retrieved. \citet{Boe_12} derive a relation giving this offset as a function of the $B-V$ color index for late-type stars (G to K). The offset $\Delta_\mathrm{E-S}$ ranges from 0 to $-0.25$\,\ensuremath{\mbox{km}\,\mbox{s}^{-1}}\ for these spectral types. 

To constrain this offset for the considered spectral type, publicly available spectra of Vega (HD~172167) are retrieved from the \'ELODIE and SOPHIE archives (respectively 16 observations and 10, see Table\,\ref{indivspectra}), and radial velocities are derived the exact same way. The respective average velocities are: $-13.44\pm 0.05$\,\ensuremath{\mbox{km}\,\mbox{s}^{-1}}\ and $-13.46\pm 0.07$\,\ensuremath{\mbox{km}\,\mbox{s}^{-1}}. The offset for the considered spectral type is then defined as the difference between these average velocities:
\begin{equation}
\Delta_\mathrm{E-S}(\mathrm{A0}) = 0.02\pm 0.09\,\ensuremath{\mbox{km}\,\mbox{s}^{-1}}.
\end{equation}
This offset is not significantly different from zero, and we choose not to correct the derived radial velocities for our targets.

\subsection{Suspected binaries}

The suspicion of binarity from the CCF is raised by the variation of the radial velocity, when several observations are available, and/or by an asymmetric shape of individual CCF. 

The variation is taken as significant when the ratio of the external error over the internal error is $E/I\gtrsim 2$ \citep{Abt_72}. The external error is chosen as the standard deviation of the measurements for the different available observations, and the internal error is the one estimated using the formulation from \citet{Zur03}, which increases with the rotational broadening. Table\,\ref{binaries_deltavr} lists the ten stars with  $E/I> 2$. The number of observations for a given star remains small, as a radial velocity follow-up was not intended. These stars are suspected single-lined spectroscopic binaries (SB1).

\begin{table}[!t]
\centering
\caption{List of targets showing a variation in their radial velocity measurements, with the number of observations ($N$), the average barycentric radial velocity $\langle$RV$\rangle$, the external error $E$  and the internal error $I$.}
\label{binaries_deltavr}
\begin{tabular}{rcrrr}
\hline
\hline
\multicolumn{1}{c}{HD} &  $N$   &\multicolumn{1}{c}{$\langle$RV$\rangle$}&\multicolumn{1}{c}{$E$} & \multicolumn{1}{c}{$I$} \\
    &        & (\ensuremath{\mbox{km}\,\mbox{s}^{-1}})             & (\ensuremath{\mbox{km}\,\mbox{s}^{-1}})&(\ensuremath{\mbox{km}\,\mbox{s}^{-1}})\\
\hline
  1561 & 2 & $ -9.2$ & 13.01 &  0.64 \\ 
 20149 & 5 & $-11.0$ &  1.33 &  0.20 \\ 
 46642 & 2 & $ 35.0$ &  4.80 &  0.73 \\ 
 72660 & 4 & $  4.2$ &  0.70 &  0.03 \\ 
119537 & 2 & $ -3.8$ & 82.92 &  0.46 \\ 
156653 & 2 & $  3.3$ &  1.32 &  0.57 \\ 
174567 & 2 & $-12.5$ &  2.36 &  0.46 \\ 
176984 & 2 & $-45.7$ &  2.78 &  0.50 \\ 
196724 & 2 & $-20.1$ &  2.59 &  0.41 \\ 
199095 & 2 & $-10.8$ & 20.14 &  0.49 \\ 

\hline
\end{tabular}
\end{table}

\begin{figure}[!t]
\centering
\resizebox{\hsize}{!}{\includegraphics{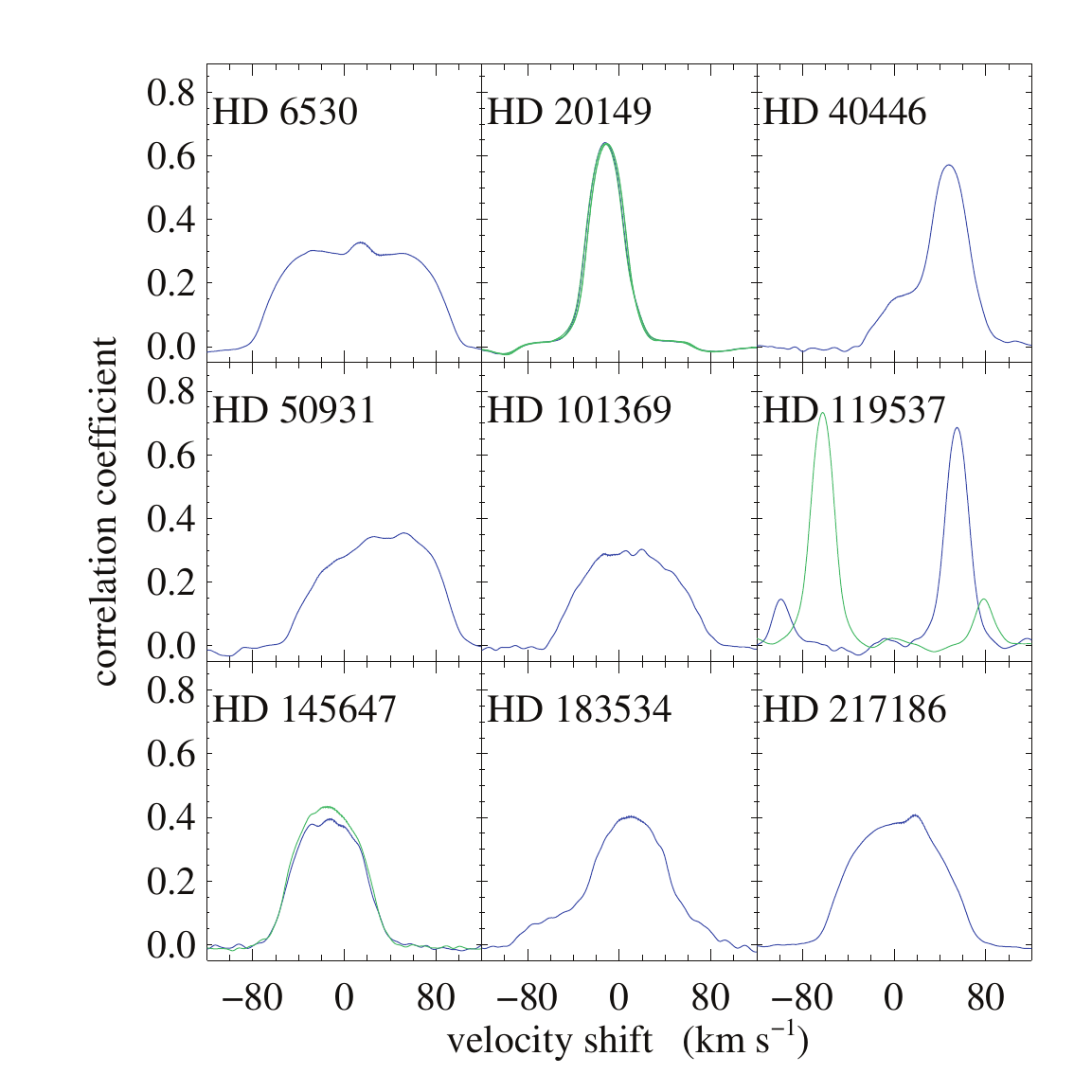}}
\caption{Cross-correlation functions of suspected binary stars from their asymmetric profiles. When available, several observations are overplotted. The velocity axis takes the barycentric correction into account.}
\label{XC}
\end{figure}

The stars that display an asymmetric CCF are shown in Fig.\,\ref{XC}. For two of them, several observations are available, and a variable radial velocity is noticed. The objects are suspected to be double-lined spectroscopic binaries (SB2).

Details on them and comparison with literature data can be found in Appendix\,\ref{comments}.

\section{Atmospheric parameters}
\label{sect_params}

We use the revised version of the \textsc{uvbybeta} code written by \cite{Nai_93} 
in order to derive effective temperatures (\ensuremath{T_\mathrm{eff}}) and surface gravities (\ensuremath{\log g}). The \textsc{uvbybetanew} relies on the calibration of the Str\"{o}mgren photometry indices $uvby\beta$ in terms of \ensuremath{T_\mathrm{eff}}\ and \ensuremath{\log g}. The photometric data are taken from \cite{HakMed98}. The derived fundamental parameters are displayed in Table\,\ref{tabstars}. According to \cite{Nai_93}, errors on effective temperature are of the order of 2\% for $\ensuremath{T_\mathrm{eff}} <10000$\,K. The accuracy of surface gravity ranges from $\approx 0.1$\,dex for early A-type stars to $\approx0.25$\,dex for hot B stars. In our study we fix the errors on \ensuremath{T_\mathrm{eff}}\ and \ensuremath{\log g}\ to be $\pm$125\,K and $\pm$0.2\,dex respectively.

\begin{figure}[!t]
\centering
\resizebox{\hsize}{!}{\includegraphics{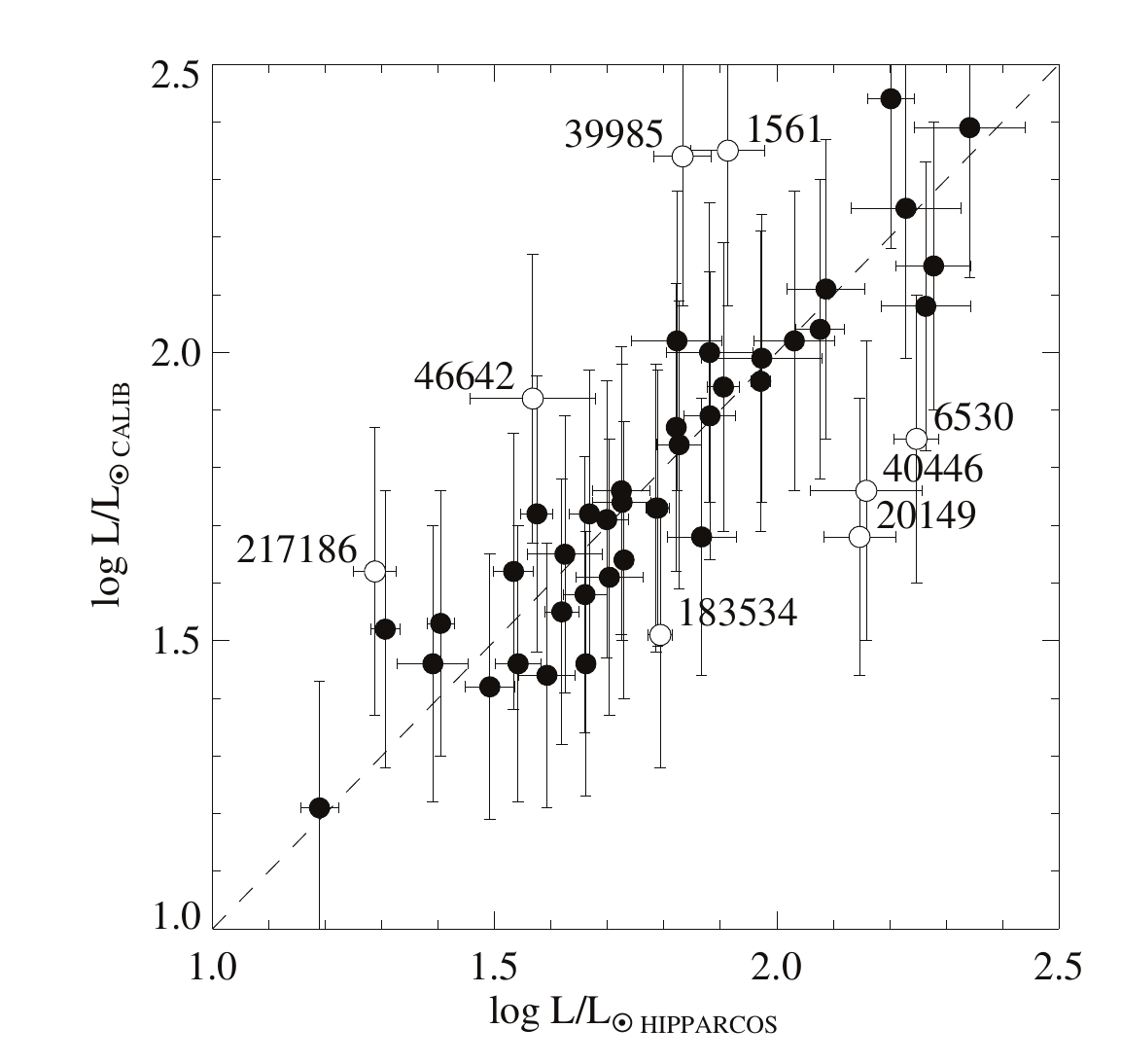}}
\caption{Consistency check using the luminosity derived from the \citet{Tos_10} calibration and the HIPPARCOS parallaxes. The dashed line is the one-to-one relation. The outliers are indicated by open symbols, together with their HD~number.}
\label{Barry}
\end{figure}

A consistency check on \ensuremath{T_\mathrm{eff}}\ and \ensuremath{\log g}\ is performed by comparing the luminosity derived from the radius calibration \citep{Tos_10}, and
the luminosity derived from HIPPARCOS parallaxes. 
\citet{Tos_10} give a polynomial expression of the stellar radius as a function of \ensuremath{T_\mathrm{eff}}, \ensuremath{\log g}\ and [Fe/H] values. The luminosity is then derived from the Stefan-Boltzmann law. Absolute magnitudes are derived from the HIPPARCOS parallaxes \citep{vLe07} and from bolometric corrections in the $V$-band, interpolated in the tables from \citet{Bel_98}. The adopted bolometric luminosity parameter $\log L/L_\odot$, given in Table\,\ref{tabstars}, is derived by adopting $M^\mathrm{bol}_\odot=4.75\,$ \citep{Aln73}. 
Figure\,\ref{Barry} compares the luminosity values. The uncertainty from the calibrated luminosity is dominated by the \ensuremath{\log g}\ uncertainty. In most cases the two agree to within the uncertainties, but a few outliers are present. These eight stars are indicated in Fig.\,\ref{Barry} and seven out of them  are suspected binaries from the previous section. The new outlier is HD~39985 (see Appendix\,\ref{comments}). In this plot, HD~33654 is out of range; the low gravity ($\ensuremath{\log g}=2.9$) derived from the photometry indicates a giant star, which is confirmed by the luminosity derived from the HIPPARCOS data: $\log L/L_\odot=3.63{\scriptstyle\pm0.84}$, far brighter than the main-sequence. It is misclassified as a class V luminosity star.

At this point, the following stars are considered as SB2 with atmospheric parameters contaminated by their multiplicity (no abundances are derived): HD~1561, HD~6530, HD~20149, HD~39985, HD~40446, HD~46642, HD~50931, HD~101369, HD~119537, HD~145647, HD~183534, HD~217186.
In addition, the following stars are considered as binaries without contamination of their atmospheric parameters: HD~72660, HD~156653, HD~174567, HD~176984, HD~196724. HD~199095.
HD~33654 is also discarded from the sample as we focus on main-sequence stars.

\begin{figure}[!t]
\centering
\resizebox{\hsize}{!}{\includegraphics{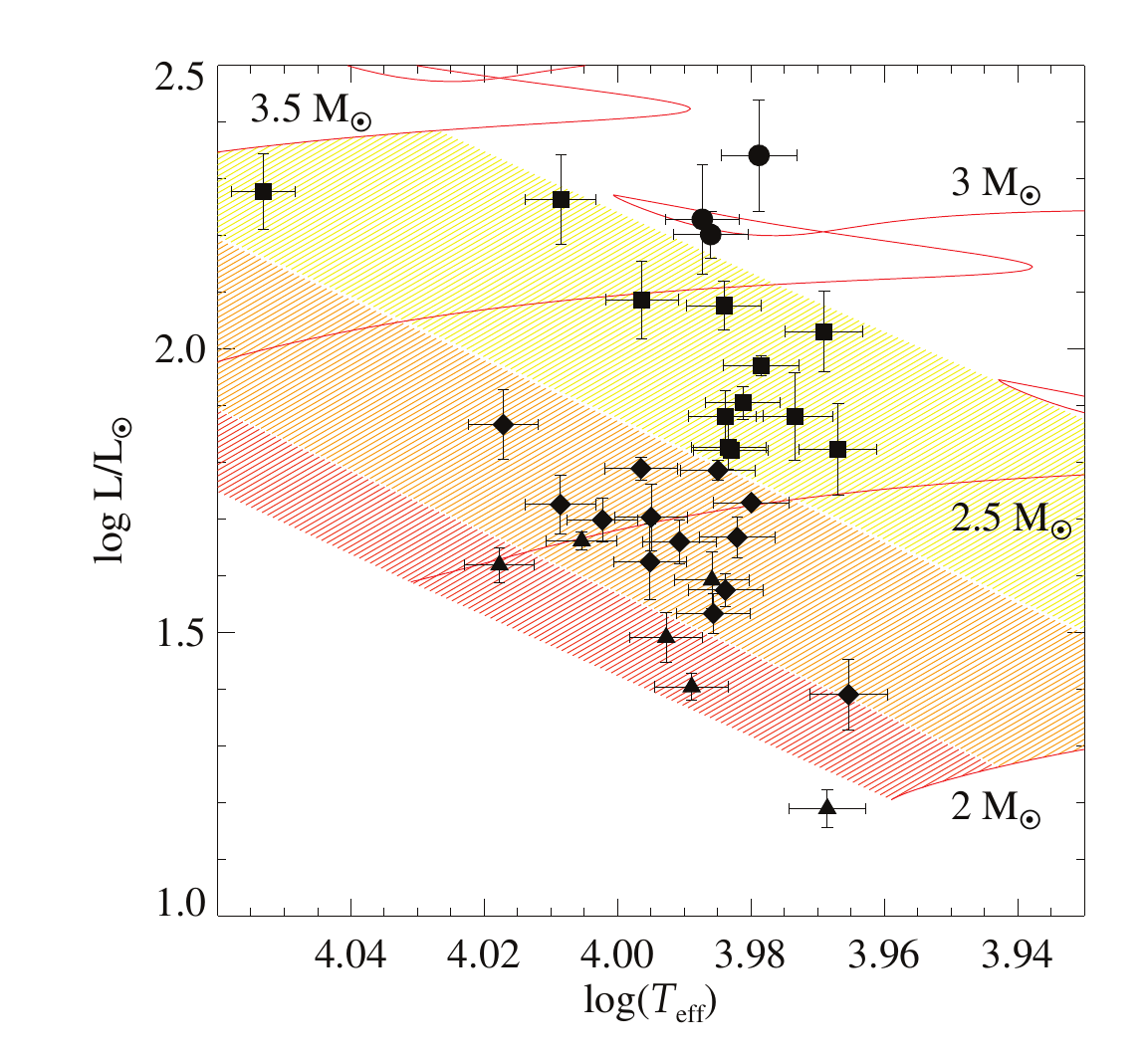}}
\caption{H-R diagram of the sample. Stars are plotted with different symbols according to their \ensuremath{\log g}, and the same limits are used to define regions from the model gravity: triangles for $4.2\le\ensuremath{\log g}$ (red), diamonds for $3.95\le\ensuremath{\log g} < 4.2$ (orange), squares for $3.65\le\ensuremath{\log g} < 3.95$ (yellow) and circles for $\ensuremath{\log g} < 3.65$. Evolutionary tracks from BaSTI 
\citep{Pii_06}, for $Z=0.0198$, with overshooting, are overplotted for the indicated stellar masses.
}
\label{HR}
\end{figure}

Figure\,\ref{HR} shows the non-SB2 stars of the sample in the H-R diagram.
Luminosities are derived from the trigonometric parallaxes.
Evolutionary tracks from BaSTI\footnote{\url{http://albione.oa-teramo.inaf.it}} \citep{Pii_06} are computed with overshooting and for a solar metallicity of $Z=0.0198$.
Considering the end of the main-sequence as its reddest point, just before the hook of the evolutionary track when the overall contraction starts, we chose three limits in $\log g$ plotted in Fig.\,\ref{HR}. The first two limits in \ensuremath{\log g}\ correspond to about one third and two thirds of lifetime on the main sequence for our mass range, according to BaSTI models. At the end of the main sequence the gravity given by BaSTI models is about $\log g=3.5$ for our mass range, and the last limit, $\log g=3.65$, corresponds to about 99\% of the lifetime on the main sequence and defines a region that does not overlap with the hook in terms of gravity and position in the H-R diagram.

These fundamental parameters are used to calculate LTE model atmospheres using ATLAS9 \citep{Kuz93}. The ATLAS9 code assumes a plane-parallel geometry, a gas in hydrostatic and radiative equilibrium, and LTE. As explained in \citet{Gen_08}, ATLAS9 model atmospheres are calculated assuming \cite{GreSal98} solar abundances and using the prescriptions given by \citet{Smy04} for convection.

\section{Rotational velocities}
\label{sect_vsini}

The \ensuremath{v\sin i}\ values used for the selection of this sample are taken from \citet{Ror_02b} who combine determinations from Fourier analysis with the catalog published by \citet{AbtMol95}. Although these values are statistically corrected from the shift between both scales, this dataset gives the opportunity to derive new and homogeneous projected rotational velocities.

\subsection{Determination of \ensuremath{v\sin i}}

The spectral synthesis described in Sect.\,\ref{sect_abund} provides determinations of \ensuremath{v\sin i}. The same Fourier analysis as in \citet{Ror_02a,Ror_02b} is also applied to provide an independent determination of \ensuremath{v\sin i}, from the position of the first zero in the Fourier transform (hereafter FT) of individual lines.

\citet{Diz_10} point out the fact that \citet{Ror_02a,Ror_02b} consider a fixed value of the linear limb-darkening coefficient, $\epsilon=0.6$, to derive \ensuremath{v\sin i}, neglecting the variation of this coefficient with \ensuremath{T_\mathrm{eff}}, \ensuremath{\log g}\ and wavelength. 
The expected variation of the limb-darkening coefficient in our sample stars can be estimated using the values of $\epsilon$ tabulated by \citet{Clt00}. In the Johnson $B$ band, $\epsilon$ varies from 0.64 to 0.56 when the effective temperature increases from 9000 to 11000\,K. 
 
In order to improve the determination of \ensuremath{v\sin i}, $\epsilon$ is taken into account by comparing the position of the first zero with a theoretical rotational profile computed with the linear limb-darkening coefficient derived from the atmospheric parameters of the stars, interpolated in the tabulated values given in \citet{Clt00}, for the $B$ band. The derived value of $\epsilon$ is given in Table\,\ref{tabstars}.

The line candidates for \ensuremath{v\sin i}\ determination are chosen among the list of 23 lines given by \citet{Ror_02b}, lying in the spectral range 4215--4577\,\AA. They are retained using criteria based on their shapes in the wavelength domain and in the frequency domain. Error on the \ensuremath{v\sin i}\ is taken as the standard deviation of single line determinations. The results are given in Table\,\ref{tabstars}.

\subsection{Rotational velocity scale}

\begin{figure}[!ht]
\centering
\resizebox{\hsize}{!}{\includegraphics[width=\textwidth,clip]{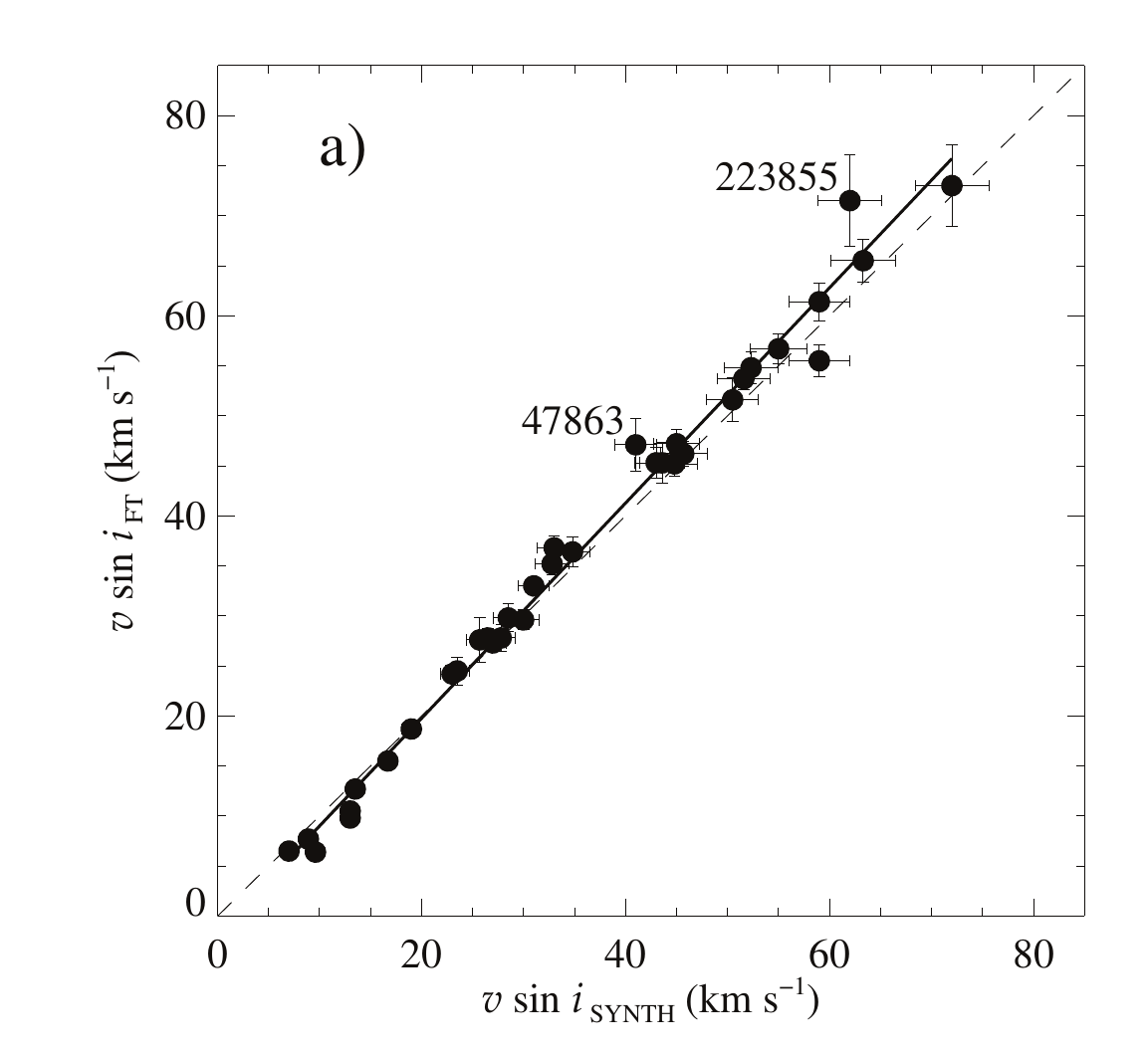}}
\resizebox{\hsize}{!}{\includegraphics[width=\textwidth,clip]{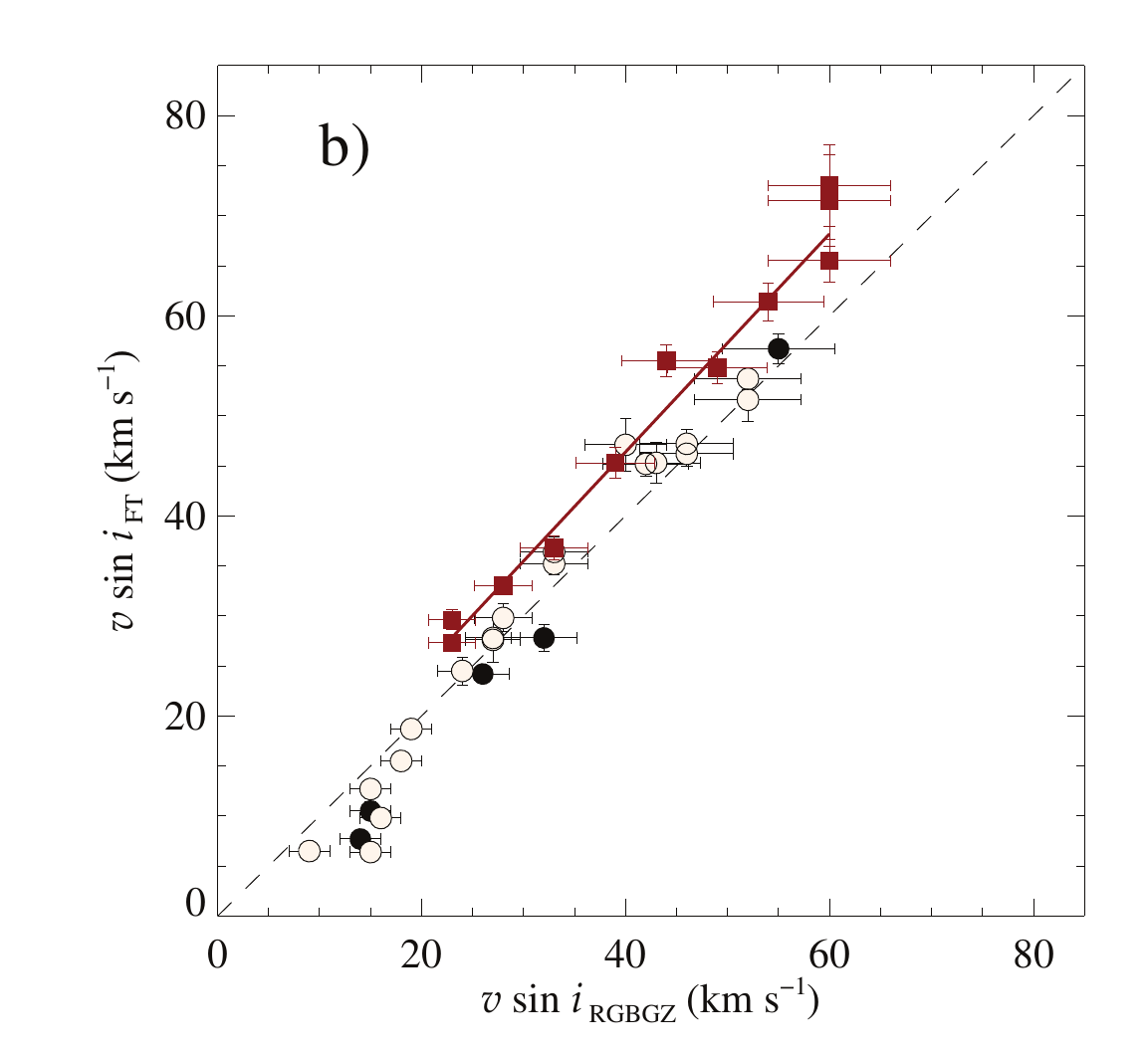}}
\caption{Comparison of the different \ensuremath{v\sin i}\ determinations given in Table\,\ref{tabstars}, excluding SB2 stars.  \textbf{(a)} The \ensuremath{v\sin i}\ derived from the first zero of the FT is compared to the one derived from the spectral synthesis (SYNTH). The linear fit is shown by the thick solid line. The two stars with the largest differences are labeled. \textbf{(b)} The \ensuremath{v\sin i}\ derived from the first zero of the FT is compared to the homogenized merged catalog from \citet[RGBGZ]{Ror_02b}. The different symbols stand from the source of the data in the latter: filled squares were from \citet[AM]{AbtMol95}, filled circles were derived by RGBGZ using Fourier analysis, open circles are a combination of both sources. The linear fit on the subsample with data from AM is represented by the thick solid line. In both panels, the one-to-one relation is represented by the dashed line.}
\label{vsinicomp}
\end{figure}

Figure\,\ref{vsinicomp}a compares the \ensuremath{v\sin i}\ derived using FT profiles in the previous paragraph and the ones resulting from the spectral synthesis (described in Sect.\,\ref{sect_abund}), from the same data. In this comparison, the suspected SB2 have been discarded. The agreement is very good and the linear relation between both scales is given by:
\begin{equation}
\ensuremath{v\sin i}_\mathrm{FT} = (1.08\pm0.01)\,\ensuremath{v\sin i}_\mathrm{SYNTH} - 1.8\pm0.3.
\end{equation}
The linear fit is overplotted on the data. The slope of the fit shows that above $\sim30\,\ensuremath{\mbox{km}\,\mbox{s}^{-1}}$, \ensuremath{v\sin i}\ derived from FT are slightly higher than the result of the spectral synthesis. This is due to the fact that the FT method uses individual line profiles and is therefore more sensitive to blends than spectral synthesis. As mentioned by \citet{Ror_02b}, effects of blends are noticeable on the individual \ensuremath{v\sin i}\ when compared with the value derived from \ion{Mg}{ii} triplet at 4481\,\AA. Two stars in Fig.\,\ref{vsinicomp}a (and Table\,\ref{tabstars}) show significant differences in \ensuremath{v\sin i}: HD~47863 and HD~223855.

Figure\,\ref{vsinicomp}b compares the new determinations to the homogenized values given by \citet[RGBGZ]{Ror_02b}. These latter determinations result from the merging of data from \citet[AM]{AbtMol95} and values derived from FT. The different symbols in the plot correspond to the combined source: AM only, FT only, or combination of both.
The systematic overestimation at low \ensuremath{v\sin i}\ in the determination from \citet{Ror_02b} results from the lower spectral resolution of \'ELODIE compared to SOPHIE in this work. 
A significant shift is noticed with the scaled values from AM, whereas no systematic effect is observed when comparing to values derived from FT. The linear relation between both scales (displayed in Fig.\,\ref{vsinicomp}b) is given by:
\begin{equation}
\ensuremath{v\sin i}_\mathrm{FT} = (1.09\pm0.03)\,\ensuremath{v\sin i}_\mathrm{RGBGZ} + 2.8\pm1.2.
\end{equation}
The scaling relation between AM and the FT results is derived by \citet{Ror_02b} using spectral types from B8 to F2. This relation could slightly vary with the spectral type, as the comparison restricted to A0--A1-type stars suggests. 
The statistical correction applied by \citet{Ror_02b} is $\ensuremath{v\sin i}=1.05\,\ensuremath{v\sin i}_\mathrm{AM}+7.5$ and the resulting values for A0--A1 stars may be undercorrected.

\section{Abundance analysis}
\label{sect_abund}

For the rotational velocity range in our sample ($\ensuremath{v\sin i}\le 65$\,\ensuremath{\mbox{km}\,\mbox{s}^{-1}}), the most appropriate method to derive individual chemical abundances is the use of spectrum synthesis technique. Specifically, we iteratively adjust LTE synthetic spectra to the observed ones by minimizing  the $\chi^2$ of the models to the observations using \citeauthor{Taa95}'s (\citeyear{Taa95}) iterative procedure \citep[see][for a detailed discussion]{Gen_08}. 
 \begin{figure*}[!htp]
\centering
\includegraphics[width=0.9\textwidth]{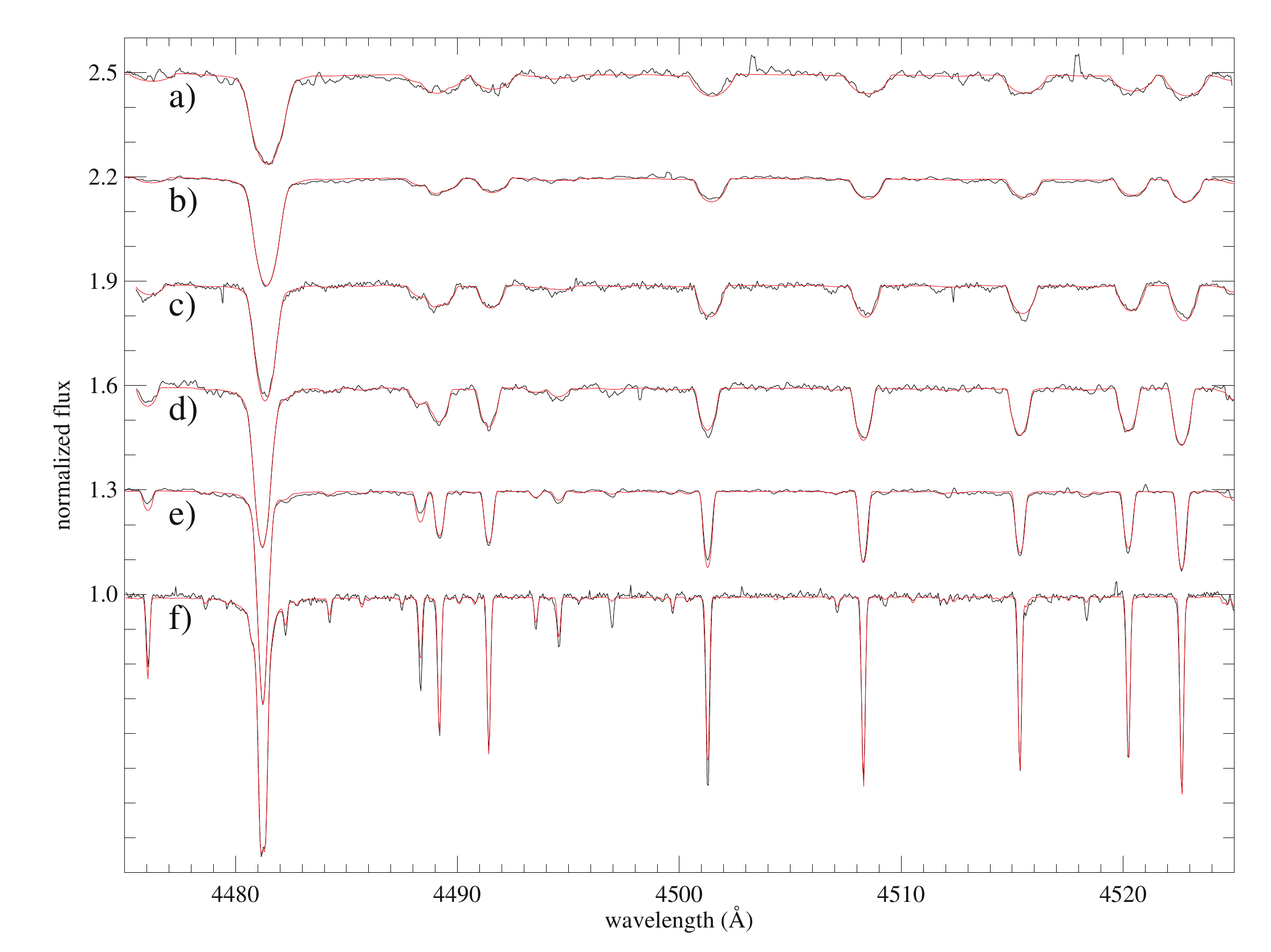}
\caption{Selected observed spectra for six A-type stars with different values of \ensuremath{v\sin i}\ (in black) shifted in flux by steps of 0.3 for clarity reasons. In red lines are the respective synthetic spectra in a region rich in \ion{Fe}{ii}, \ion{Ti}{ii}, and \ion{Mg}{ii} lines. The spectra are sorted from top to bottom by decreasing \ensuremath{v\sin i}. The stars are \textbf{(a)} \object{HD~101369}, \textbf{(b)} \object{HD~133962}, \textbf{(c)} \object{HD~107655}, \textbf{(d)} \object{HD~65900}, \textbf{(e)} \object{HD~58142} and \textbf{(f)} \object{HD~72660}. Spectra are shifted to rest wavelengths.}
\label{ObsSS}
\end{figure*}

\subsection{Spectrum synthesis}

\onllongtab{
\begin{sidewaystable*}
\caption{Abundances for the 34 main-sequence targets not identified as contaminated. Ellipsis dots (...) stand for abundances that could not be determined for the corresponding object.}
\label{abundances}
{\scriptsize
\setlength\tabcolsep{2pt}
\begin{tabular}{rrrrrrrrrrrrrrrrr}
\hline\hline
  \multicolumn{1}{c}{HD} &
  \multicolumn{1}{c}{[C/H]} &
  \multicolumn{1}{c}{[O/H]} &
  \multicolumn{1}{c}{[Mg/H]} &
  \multicolumn{1}{c}{[Si/H]} &
  \multicolumn{1}{c}{[Ca/H]} &
  \multicolumn{1}{c}{[Sc/H]} &
  \multicolumn{1}{c}{[Ti/H]} &
  \multicolumn{1}{c}{[Cr/H]} &
  \multicolumn{1}{c}{[Fe/H]} &
  \multicolumn{1}{c}{[Ni/H]} &
  \multicolumn{1}{c}{[Sr/H]} &
  \multicolumn{1}{c}{[Y/H]} &
  \multicolumn{1}{c}{[Zr/H]} &
  \multicolumn{1}{c}{[Ba/H]} \\
\hline
  1439 & \multicolumn{1}{c}{...} & $0.018\pm0.018$ & $0.220\pm0.140$ & $0.231\pm0.061$ & $-0.232\pm0.118$ & $-0.098\pm0.125$ & $-0.023\pm0.048$ & $0.265\pm0.135$ & $0.072\pm0.050$ & \multicolumn{1}{c}{...} & $0.630\pm0.150$ & $0.685\pm0.215$ & $0.575\pm0.004$ & \multicolumn{1}{c}{...} \\
 21050 & $-0.360\pm0.080$ & $0.023\pm0.041$ & $0.170\pm0.190$ & $0.175\pm0.010$ & $-0.133\pm0.138$ & $-0.130\pm0.237$ & $-0.166\pm0.056$ & $0.278\pm0.160$ & $-0.075\pm0.190$ & \multicolumn{1}{c}{...} & $-0.350\pm0.150$ & \multicolumn{1}{c}{...} & $0.200\pm0.185$ & \multicolumn{1}{c}{...} \\
 25175 & $-0.365\pm0.105$ & $-0.010\pm0.030$ & $0.025\pm0.145$ & $0.180\pm0.103$ & $-0.115\pm0.035$ & $0.119\pm0.104$ & $-0.151\pm0.092$ & $0.118\pm0.058$ & $-0.160\pm0.181$ & $0.021\pm0.311$ & $-0.050\pm0.150$ & $0.046\pm0.150$ & $0.733\pm0.077$ & $0.817\pm0.150$ \\
 28780 & $-0.528\pm0.047$ & $-0.235\pm0.065$ & $-0.070\pm0.190$ & $-0.105\pm0.175$ & $-0.007\pm0.234$ & $-0.083\pm0.068$ & $0.013\pm0.058$ & $0.063\pm0.098$ & $-0.050\pm0.088$ & $0.280\pm0.143$ & $0.310\pm0.150$ & $0.225\pm0.045$ & $0.670\pm0.116$ & $0.744\pm0.069$ \\
 30085 & $-0.518\pm0.061$ & $-0.165\pm0.015$ & $-0.295\pm0.092$ & $0.056\pm0.033$ & $0.003\pm0.053$ & $-0.193\pm0.179$ & $0.163\pm0.171$ & $0.816\pm0.357$ & $-0.067\pm0.155$ & \multicolumn{1}{c}{...} & $1.118\pm0.150$ & $2.096\pm0.024$ & $2.062\pm0.082$ & $1.175\pm0.035$ \\
 47863 & $-0.185\pm0.025$ & $0.112\pm0.150$ & $0.153\pm0.127$ & $0.200\pm0.000$ & $0.310\pm0.021$ & $-0.057\pm0.187$ & $-0.432\pm0.209$ & $-0.014\pm0.099$ & $-0.270\pm0.519$ & $-0.039\pm0.041$ & $-0.758\pm0.150$ & $-0.128\pm0.150$ & $0.466\pm0.024$ & $0.830\pm0.150$ \\
 58142 & $-0.452\pm0.056$ & $-0.246\pm0.056$ & $-0.065\pm0.175$ & $0.269\pm0.221$ & $-0.057\pm0.417$ & $-0.466\pm0.118$ & $-0.165\pm0.085$ & $0.055\pm0.078$ & $-0.004\pm0.124$ & $0.287\pm0.136$ & $0.410\pm0.150$ & $0.313\pm0.105$ & $0.550\pm0.097$ & $0.912\pm0.168$ \\
 65900 & $-0.564\pm0.379$ & $-0.425\pm0.058$ & $0.073\pm0.103$ & $0.421\pm0.223$ & $0.063\pm0.243$ & $-0.298\pm0.216$ & $-0.029\pm0.093$ & $0.378\pm0.076$ & $0.247\pm0.121$ & $0.674\pm0.093$ & $0.805\pm0.160$ & $0.837\pm0.097$ & $1.123\pm0.166$ & $1.433\pm0.150$ \\
 67959 & $-0.403\pm0.026$ & $-0.270\pm0.047$ & $0.025\pm0.245$ & \multicolumn{1}{c}{...} & $-0.079\pm0.015$ & $-0.225\pm0.167$ & $0.153\pm0.088$ & $0.238\pm0.068$ & $0.190\pm0.075$ & $0.494\pm0.115$ & $0.786\pm0.150$ & \multicolumn{1}{c}{...} & \multicolumn{1}{c}{...} & \multicolumn{1}{c}{...} \\
 72660 & $-0.610\pm0.082$ & $-0.506\pm0.066$ & $0.094\pm0.127$ & $0.295\pm0.025$ & $0.053\pm0.110$ & $-0.245\pm0.344$ & $0.459\pm0.114$ & $0.800\pm0.306$ & $0.456\pm0.061$ & $0.696\pm0.105$ & $1.220\pm0.150$ & $0.835\pm0.092$ & $1.018\pm0.051$ & \multicolumn{1}{c}{...} \\
 73316 & $-0.308\pm0.056$ & $0.054\pm0.050$ & $0.056\pm0.146$ & $0.068\pm0.034$ & $0.036\pm0.200$ & $-0.374\pm0.352$ & $-0.072\pm0.109$ & $0.085\pm0.082$ & $-0.210\pm0.132$ & \multicolumn{1}{c}{...} & $-0.434\pm0.150$ & \multicolumn{1}{c}{...} & $0.430\pm0.150$ & $-0.096\pm0.082$ \\
 83373 & \multicolumn{1}{c}{...} & $-0.235\pm0.005$ & $0.095\pm0.165$ & $0.407\pm0.055$ & $-0.065\pm0.025$ & $0.073\pm0.232$ & $-0.166\pm0.067$ & $0.464\pm0.168$ & $0.099\pm0.170$ & \multicolumn{1}{c}{...} & $0.341\pm0.150$ & $0.886\pm0.150$ & $0.640\pm0.120$ & $1.110\pm0.150$ \\
 85504 & $0.251\pm0.146$ & $0.240\pm0.076$ & $0.412\pm0.252$ & $0.867\pm0.351$ & $0.129\pm0.119$ & $0.358\pm0.234$ & $-0.037\pm0.088$ & $0.382\pm0.167$ & $0.033\pm0.111$ & $0.260\pm0.270$ & $-0.238\pm0.150$ & $0.532\pm0.172$ & $0.613\pm0.448$ & \multicolumn{1}{c}{...} \\
 89774 & $-0.363\pm0.027$ & $0.167\pm0.096$ & $0.030\pm0.204$ & $0.149\pm0.111$ & $-0.097\pm0.187$ & $-0.269\pm0.228$ & $-0.115\pm0.069$ & $0.165\pm0.073$ & $-0.038\pm0.149$ & $0.057\pm0.094$ & $-0.099\pm0.150$ & $0.474\pm0.013$ & $0.919\pm0.019$ & $1.255\pm0.385$ \\
 95418 & $-0.510\pm0.010$ & $-0.327\pm0.024$ & $0.052\pm0.128$ & $0.438\pm0.273$ & $0.203\pm0.195$ & $-0.115\pm0.107$ & $-0.017\pm0.041$ & $0.318\pm0.095$ & $0.249\pm0.110$ & $0.400\pm0.081$ & $0.839\pm0.156$ & $0.750\pm0.115$ & $1.057\pm0.143$ & \multicolumn{1}{c}{...} \\
104181 & $-0.040\pm0.150$ & $0.028\pm0.007$ & $0.090\pm0.150$ & $0.346\pm0.317$ & $-0.020\pm0.150$ & $0.200\pm0.150$ & $0.035\pm0.175$ & $-0.008\pm0.102$ & $-0.207\pm0.025$ & $-0.062\pm0.399$ & $-0.336\pm0.150$ & \multicolumn{1}{c}{...} & $0.750\pm0.261$ & \multicolumn{1}{c}{...} \\
107655 & $-0.750\pm0.150$ & $-0.582\pm0.150$ & $-0.117\pm0.150$ & $0.039\pm0.073$ & \multicolumn{1}{c}{...} & $0.178\pm0.066$ & $-0.035\pm0.116$ & $0.559\pm0.089$ & $0.076\pm0.055$ & $0.771\pm0.230$ & $0.197\pm0.150$ & $0.798\pm0.149$ & $0.815\pm0.176$ & $0.715\pm0.150$ \\
127304 & $-0.667\pm0.029$ & $-0.414\pm0.050$ & $0.013\pm0.147$ & $0.605\pm0.316$ & $-0.237\pm0.161$ & $-0.014\pm0.150$ & $0.070\pm0.027$ & $0.547\pm0.174$ & $0.233\pm0.061$ & $0.655\pm0.166$ & $0.780\pm0.150$ & \multicolumn{1}{c}{...} & $0.807\pm0.081$ & \multicolumn{1}{c}{...} \\
132145 & $-0.180\pm0.088$ & $-0.145\pm0.083$ & $0.107\pm0.183$ & $0.271\pm0.221$ & $-0.068\pm0.041$ & $-0.130\pm0.164$ & $-0.076\pm0.048$ & $0.123\pm0.049$ & $-0.097\pm0.126$ & $-0.045\pm0.174$ & $-0.490\pm0.150$ & \multicolumn{1}{c}{...} & $0.254\pm0.224$ & \multicolumn{1}{c}{...} \\
133962 & $-0.165\pm0.033$ & $-0.011\pm0.020$ & $0.005\pm0.205$ & $-0.115\pm0.150$ & $0.043\pm0.037$ & $0.000\pm0.220$ & $0.009\pm0.079$ & $0.112\pm0.069$ & $0.019\pm0.185$ & $0.422\pm0.127$ & $0.230\pm0.150$ & \multicolumn{1}{c}{...} & $0.520\pm0.270$ & \multicolumn{1}{c}{...} \\
145788 & $-0.312\pm0.165$ & $0.048\pm0.080$ & $0.230\pm0.140$ & $0.265\pm0.055$ & $-0.165\pm0.059$ & $-0.310\pm0.070$ & $-0.145\pm0.015$ & $0.142\pm0.082$ & $-0.137\pm0.149$ & $0.120\pm0.213$ & $-0.430\pm0.150$ & $0.034\pm0.085$ & $0.180\pm0.150$ & \multicolumn{1}{c}{...} \\
154228 & $-0.333\pm0.385$ & $-0.240\pm0.150$ & $0.060\pm0.172$ & $0.280\pm0.030$ & $-0.037\pm0.504$ & $-0.387\pm0.128$ & $-0.006\pm0.060$ & $0.433\pm0.134$ & $0.191\pm0.074$ & $0.440\pm0.177$ & $0.460\pm0.150$ & \multicolumn{1}{c}{...} & $0.763\pm0.214$ & \multicolumn{1}{c}{...} \\
156653 & $-0.850\pm0.110$ & $-0.292\pm0.030$ & $-0.050\pm0.280$ & $0.008\pm0.230$ & $-0.175\pm0.558$ & $0.082\pm0.072$ & $-0.190\pm0.057$ & $0.133\pm0.121$ & $0.006\pm0.102$ & $0.037\pm0.203$ & $0.245\pm0.150$ & \multicolumn{1}{c}{...} & \multicolumn{1}{c}{...} & $0.803\pm0.163$ \\
158716 & $-0.437\pm0.136$ & $-0.427\pm0.101$ & $0.055\pm0.115$ & $0.311\pm0.007$ & $-0.251\pm0.387$ & $-1.114\pm0.150$ & $0.192\pm0.025$ & $0.198\pm0.013$ & $0.235\pm0.060$ & $0.350\pm0.113$ & $0.490\pm0.150$ & $0.775\pm0.035$ & $0.755\pm0.125$ & \multicolumn{1}{c}{...} \\
172167 & $-0.210\pm0.154$ & $-0.056\pm0.115$ & $-0.266\pm0.176$ & $-0.313\pm0.184$ & $-0.315\pm0.548$ & $-0.291\pm0.199$ & $-0.225\pm0.297$ & $-0.345\pm0.144$ & $-0.410\pm0.096$ & $0.047\pm0.216$ & $-0.845\pm0.150$ & \multicolumn{1}{c}{...} & $0.160\pm0.170$ & \multicolumn{1}{c}{...} \\
174567 & $-0.450\pm0.150$ & $-0.085\pm0.131$ & $0.170\pm0.250$ & $0.195\pm0.045$ & $0.026\pm0.217$ & $-0.210\pm0.150$ & $-0.205\pm0.055$ & $0.190\pm0.190$ & $-0.036\pm0.085$ & $0.130\pm0.150$ & $-0.212\pm0.150$ & $0.218\pm0.162$ & $0.525\pm0.065$ & \multicolumn{1}{c}{...} \\
176984 & $-0.133\pm0.210$ & $-0.205\pm0.095$ & $0.260\pm0.230$ & $0.140\pm0.150$ & $0.199\pm0.226$ & $-0.110\pm0.150$ & $-0.117\pm0.043$ & $0.263\pm0.118$ & $-0.276\pm0.090$ & $0.140\pm0.450$ & $0.240\pm0.150$ & $0.145\pm0.205$ & $0.043\pm0.065$ & \multicolumn{1}{c}{...} \\
196724 & \multicolumn{1}{c}{...} & $-0.328\pm0.087$ & $0.080\pm0.150$ & $-0.030\pm0.100$ & $-0.186\pm0.017$ & $-0.214\pm0.150$ & $-0.314\pm0.100$ & $0.340\pm0.224$ & $0.008\pm0.077$ & $0.138\pm0.150$ & $0.198\pm0.150$ & $0.418\pm0.174$ & $0.585\pm0.167$ & \multicolumn{1}{c}{...} \\
198552 & $-0.198\pm0.082$ & $-0.372\pm0.032$ & $-0.120\pm0.139$ & $0.095\pm0.277$ & $0.578\pm0.392$ & $-0.470\pm0.214$ & $-0.280\pm0.035$ & $0.039\pm0.118$ & $-0.069\pm0.110$ & $0.425\pm0.122$ & $-0.120\pm0.150$ & $0.150\pm0.083$ & $0.426\pm0.092$ & \multicolumn{1}{c}{...} \\
199095 & $-0.177\pm0.108$ & $0.010\pm0.073$ & $-0.070\pm0.314$ & $0.105\pm0.162$ & $-0.043\pm0.107$ & $-0.185\pm0.176$ & $0.193\pm0.160$ & $0.089\pm0.116$ & $0.029\pm0.078$ & $0.330\pm0.344$ & $-0.040\pm0.150$ & \multicolumn{1}{c}{...} & $0.500\pm0.040$ & \multicolumn{1}{c}{...} \\
219290 & $0.160\pm0.150$ & $-0.080\pm0.057$ & $0.100\pm0.130$ & $0.106\pm0.032$ & $-0.160\pm0.151$ & $-0.017\pm0.048$ & $-0.025\pm0.070$ & $0.278\pm0.104$ & $0.015\pm0.038$ & $0.640\pm0.150$ & $0.540\pm0.150$ & \multicolumn{1}{c}{...} & $0.503\pm0.150$ & \multicolumn{1}{c}{...} \\
219485 & $-0.220\pm0.150$ & $-0.130\pm0.021$ & $0.140\pm0.150$ & $0.242\pm0.352$ & $-0.183\pm0.050$ & $-0.393\pm0.111$ & $-0.035\pm0.045$ & $0.232\pm0.242$ & $0.034\pm0.099$ & \multicolumn{1}{c}{...} & $0.550\pm0.150$ & $0.260\pm0.113$ & $0.500\pm0.150$ & \multicolumn{1}{c}{...} \\
223386 & \multicolumn{1}{c}{...} & $-0.174\pm0.081$ & $0.126\pm0.150$ & $0.052\pm0.126$ & \multicolumn{1}{c}{...} & $-0.623\pm0.257$ & $-0.315\pm0.065$ & $0.130\pm0.099$ & $-0.227\pm0.061$ & \multicolumn{1}{c}{...} & $-0.300\pm0.150$ & $0.049\pm0.051$ & $-0.020\pm0.150$ & \multicolumn{1}{c}{...} \\
223855 & $-0.163\pm0.097$ & $0.013\pm0.090$ & $0.270\pm0.150$ & $0.323\pm0.386$ & \multicolumn{1}{c}{...} & $-0.115\pm0.085$ & $-0.213\pm0.042$ & $0.022\pm0.079$ & $-0.036\pm0.224$ & \multicolumn{1}{c}{...} & $-0.730\pm0.150$ & $0.145\pm0.105$ & $0.340\pm0.150$ & \multicolumn{1}{c}{...} \\

\hline
\end{tabular}
}
\end{sidewaystable*}
}

\citeauthor{Taa95}'s procedure is divided in two parts. The first part is a modified version of \citet{Kuz92} \textsc{Width9} code and computes the opacity data. The second part of the routine computes the synthetic spectrum and minimizes the dispersion between the normalized synthetic spectrum and the observed one. 

The line list used for spectral synthesis is the one used in \citet{Gen_08,Gen_10}. All transitions between 3000 and 7000\,\AA\ from Kurucz's gfall.dat\footnote{\url{http://kurucz.harvard.edu/LINELISTS/GFALL}} line list are selected for the calculation of the synthetic spectra. The abundance determination relies mainly on unblended transitions for about 14 chemical elements (C, O, Mg, Si, Ca, Sc, Ti, Cr, Fe, Ni, Sr, Y, Zr, and Ba). Most of these lines are weak because they are formed deep in the atmosphere. They are well suited to abundance determinations as LTE should prevail in the deeper layers of the atmospheres.

The accuracy of the atomic parameters (wavelengths, lower excitation potential, oscillator strength and damping constants) is checked against more accurate and/or more recent laboratory determinations, using the VALD\footnote{\url{http://www.astro.uu.se/~vald/php/vald.php}} \citep{Kua_99} and the NIST\footnote{\url{http://physics.nist.gov/PhysRefData/ASD/lines_form.html}} databases. 
The adopted atomic data for each elements are collected in Table~\ref{linelist}, where for each
element, the wavelength, the oscillator strength, and the reference are given.

A byproduct of this procedure is the derivation of the rotational (\ensuremath{v\sin i}) and the microturbulent ($\xi_\mathrm{t}$) velocities. We first derive the rotational and microturbulent velocities using several weak and moderately strong unblended \ion{Fe}{ii} lines located between 4491.405\,\AA\ and 4522.634\,\AA\ and the \ion{Mg}{ii} triplet around 4481\,\AA\ by allowing small variations around solar abundances of Mg and Fe as explained in Sect.\,3.2.1 of \cite{Gen_08}.
The weak iron lines are very sensitive to rotational velocity but not to microturbulent velocity while the moderately strong \ion{Fe}{ii} lines are affected mostly by changes of microturbulent velocity. The \ion{Mg}{ii} triplet is sensitive to both $\xi_\mathrm{t}$ and \ensuremath{v\sin i}. The derived rotational and microturbulent velocities are displayed in Table\,\ref{tabstars}. The error on $\xi_\mathrm{t}$ is $\pm0.5$\,\ensuremath{\mbox{km}\,\mbox{s}^{-1}}\ \citep{Gen_10}. By testing the effect of the variation of the \ensuremath{v\sin i} \ on the abundance derived from the \ion{Mg}{ii} triplet, we are able to determine $\Delta (\ensuremath{v\sin i})$ that causes a variation of the abundance of about the error level ($\Delta [\mathrm{Mg/H}] \sim \sigma_\mathrm{Mg}$). On average we find that the rotational velocities have a precision estimated as 5\% of the nominal \ensuremath{v\sin i}\ and are in good agreement with those derived using the Fourier transforms (Sect.\,\ref{sect_vsini}).

Once the rotational and microturbulent velocities are fixed, we then derive the abundance that minimized the $\chi^2$ for each transition of a given chemical element. 
Figure\,\ref{ObsSS} displays the observed spectra of six A stars, for different \ensuremath{v\sin i}, with their respective best fit synthetic spectra.

\subsection{Resulting abundances}
The derived abundances are mean values and given in solar scale in Table\,\ref{abundances}, available online. For a given chemical element X, the abundance $\left[\mathrm{X/H}\right]$ is equal to the difference between the absolute abundance in the star $\log\left(\mathrm{X/H}\right)_{\star}$ and the abundance in the Sun $\log\left(\mathrm{X/H}\right)_{\odot}$ derived from \cite{GreSal98}. As done in \cite{Gen_08}, errors on the elemental abundances are estimated by the standard deviation, assuming a Gaussian distribution of the abundances derived from each line. When only one line is measured, we adopted an average error on abundances derived from the results of \citet{Gen_08,Gen_10} and \citet{GenMor08} for stars with \ensuremath{v\sin i}\ $<$ 70 \ensuremath{\mbox{km}\,\mbox{s}^{-1}}. This error is found to be $\approx$ 0.15\,dex.

\section{Cluster analysis and classification}
\label{sect_classif}
A quick look on the abundances listed in Table\,\ref{abundances} and how these data are distributed in the Sr--Sc plane (Fig.\,\ref{dendogram}b) suggests that several stars present CP characteristics. 
The size of our sample allows the use of statistical tools to disentangle the CP and normal stars and perform this classification in an automatic way, from the full abundance data set.

\subsection{Classification criteria}
 In our temperature domain, the expected CP stars are the metallic Am stars (CP1) and the magnetic Ap stars (CP2). The classical definition of CP1 stars relies on Ca, Sc, iron-peak elements and heavy elements \citep{Coi70,Prn74}, and for the CP2 stars it is based on Si, Cr, Sr and Eu \citep{Prn74}. Among the 14 species studied in this work, 10 species correspond to these classical definitions, i.e Si, Ca, Sc, Cr, Fe, Ni, Sr, Y, Zr and Ba.

\subsection{Hierarchical cluster analysis}
\begin{figure}[!t]
\centering
\resizebox{\hsize}{!}{\includegraphics[width=\textwidth,clip,viewport=135 165 680 430]{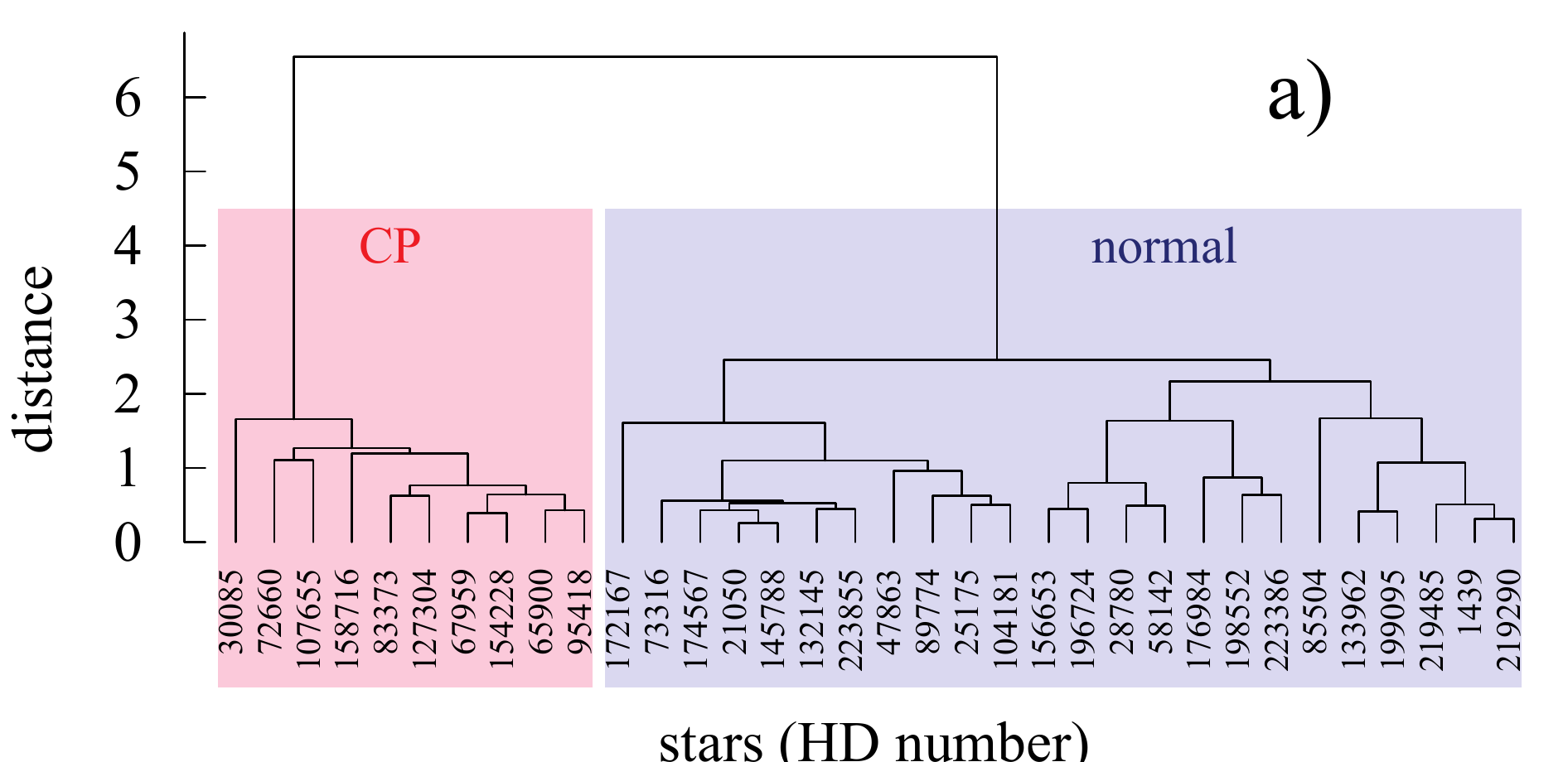}}

\resizebox{\hsize}{!}{\includegraphics{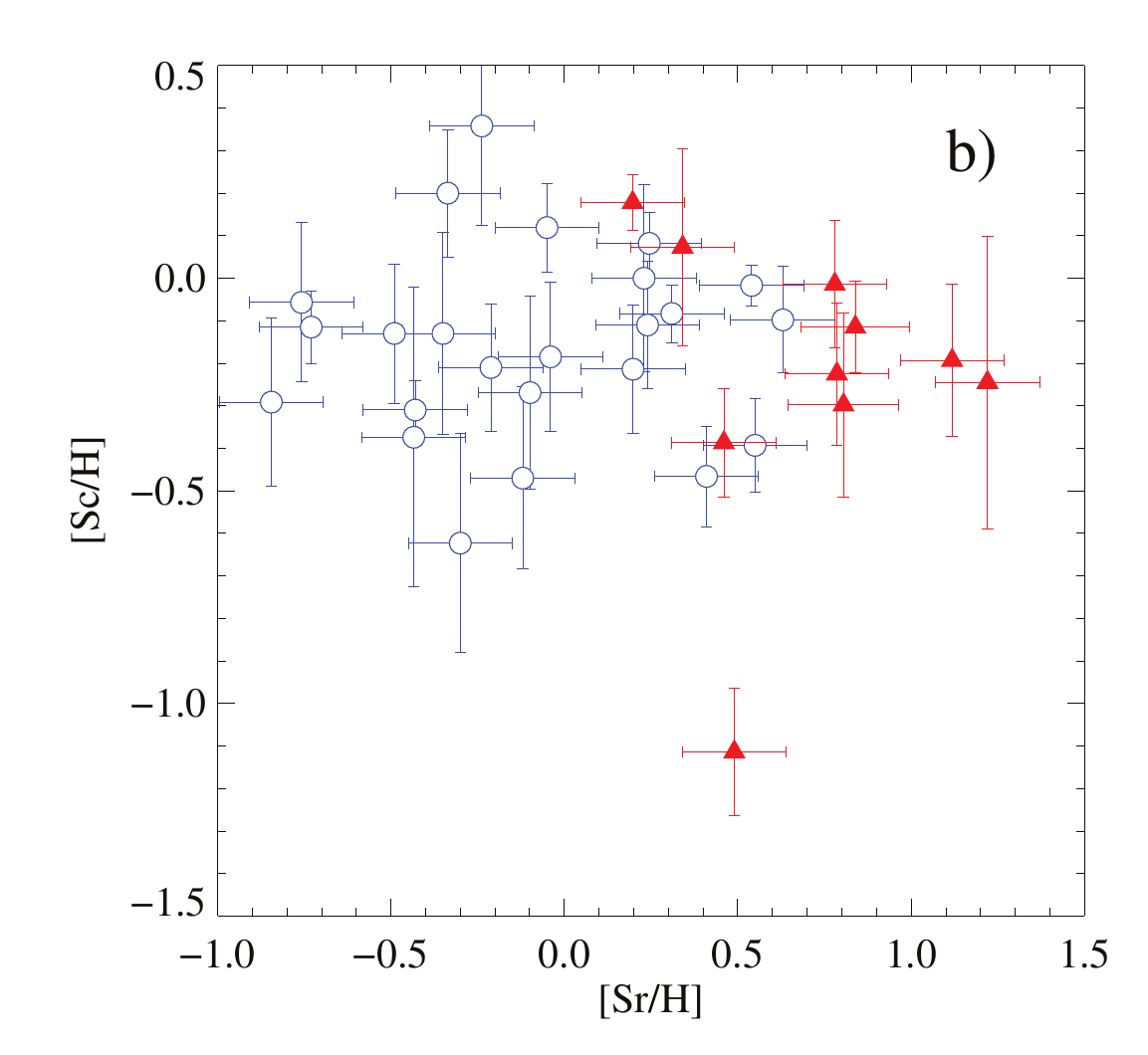}}
\caption{\textbf{(a)} Dendogram plot of the hierarchical tree resulting from the cluster analysis of the 14-species chemical abundances of the sample without errors. The x-axis gives the HD~numbers of the stars and the y-axis represents the Euclidean distance, in the normalized abundance space, between subgroups. The two main groups are identified by the labeled boxes. \textbf{(b)} Scandium abundances as a function of strontium. The two main groups identified above are projected with different symbol: filled triangle stand for CP stars, open circles represent normal stars.}
\label{dendogram}
\end{figure}

Hierarchical cluster analysis, applied to chemical abundances by \citet{CoyBod04}, is a bottom-up classification which consists in grouping data by proximity in a given space, producing a classification tree from single elements to the entire sample. It identifies clusters as a function of distance between elements. This method is applied to our data and clusters are searched for in the multi-dimensional space defined by the abundances from Table\,\ref{abundances}. Abundances are normalized so that the variation of a given elemental abundance over the full sample lies in the interval [$0,1$], ensuring that the different elements are equally weighted. Proximity is based on the Euclidean distance in the multi-dimensional normalized space. In the resulting classification tree (Fig.\,\ref{dendogram}a), two main groups appear. They are identified as CP and normal stars, based on the median [Sr/H] abundance: this element being very discriminant (Fig.\,\ref{dendogram}b), the group showing the highest median [Sr/H] abundance is associated with CP stars, and the other one with normal stars.

The construction of the classification tree does not take errors into account. In order to test the effect of the errors on the resulting memberships, new abundances are randomly simulated by adding Gaussian noise to the measured abundances, using the related standard deviation. Over 5000 simulations, targets are allocated to one group or the other, and the proportion of allocation to one group is taken as the final membership probability. 
This defines a new classification, taking the abundance errors into account. 

 This multivariate statistical analysis is performed using R\footnote{R is a language and environment for statistical computing and graphics, available at \url{http://www.r-project.org}} \citep{R}.

\subsection{Resulting classification and comparison with literature data}

\begin{figure}[!t]
\centering
\resizebox{\hsize}{!}{\includegraphics[width=\textwidth,clip]{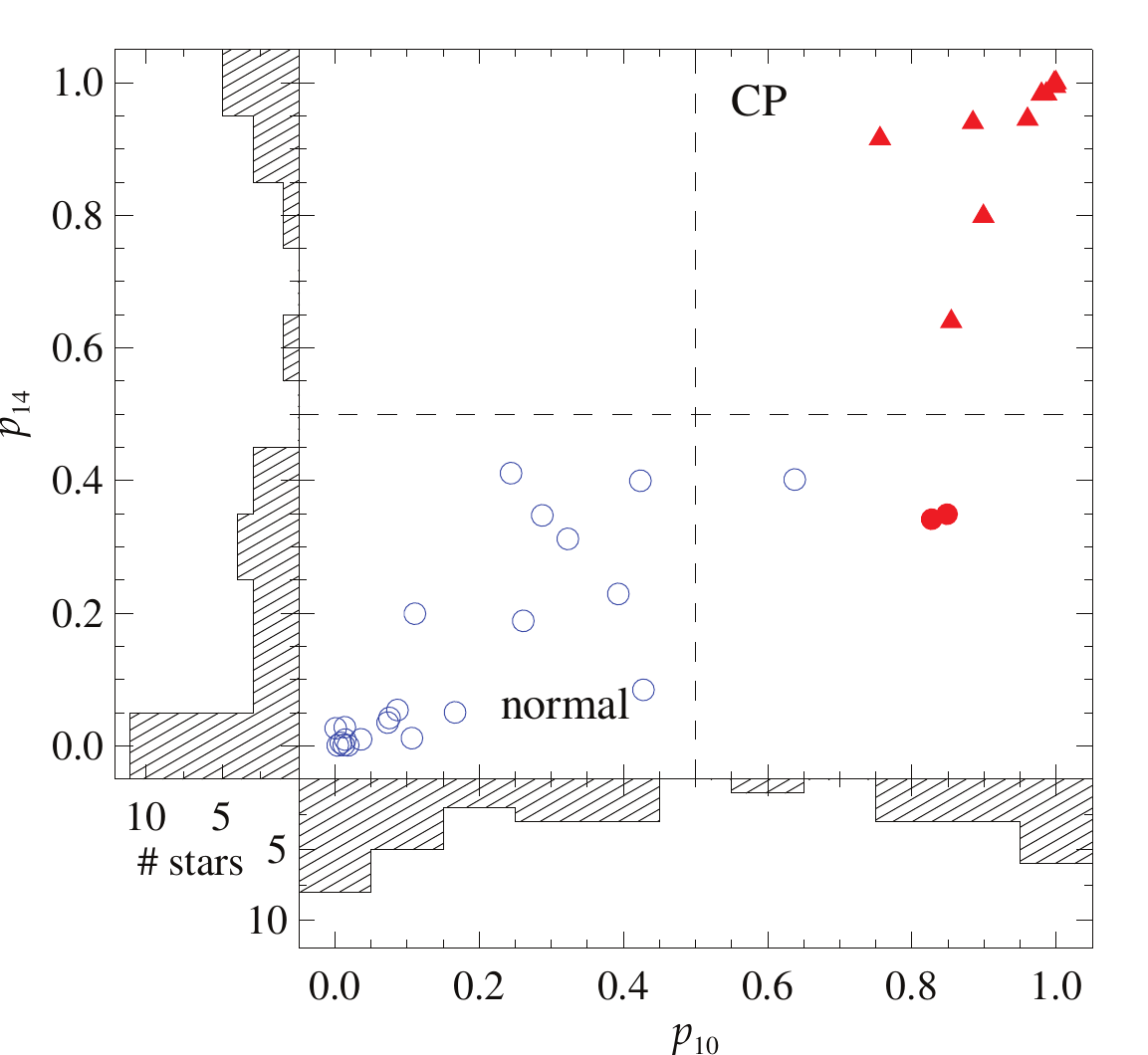}}
\caption{Membership to the CP group from the four criteria: probabilities using 10 and 14 species (respectively $p_{10}$ and $p_{14}$), direct classifications using 10 species (filled symbol for CP stars, open symbol for normal stars), and using 14 species (triangle for CP stars, circle for normal stars). Lower and left panels are the projected distributions of $p_{10}$ and $p_{14}$ respectively.}
\label{classif}
\end{figure}

Hierarchical cluster analysis is applied to the 14 species and produces the membership flag $f_{14}$ (1 for CP, 0 for normal) from the direct classification, for each star. The simulations taking errors into account give the membership probability $p_{14}$. Respectively, $f_{10}$ and $p_{10}$ are produced when applying hierarchical cluster analysis to the 10 ``classical'' species. Figure\,\ref{classif} displays these four criteria, also listed in Table\,\ref{table_classif} (available online).

\begin{figure}[!th]
\centering
\resizebox{\hsize}{!}{\includegraphics[width=\textwidth,viewport= 140 25 700 565,clip]{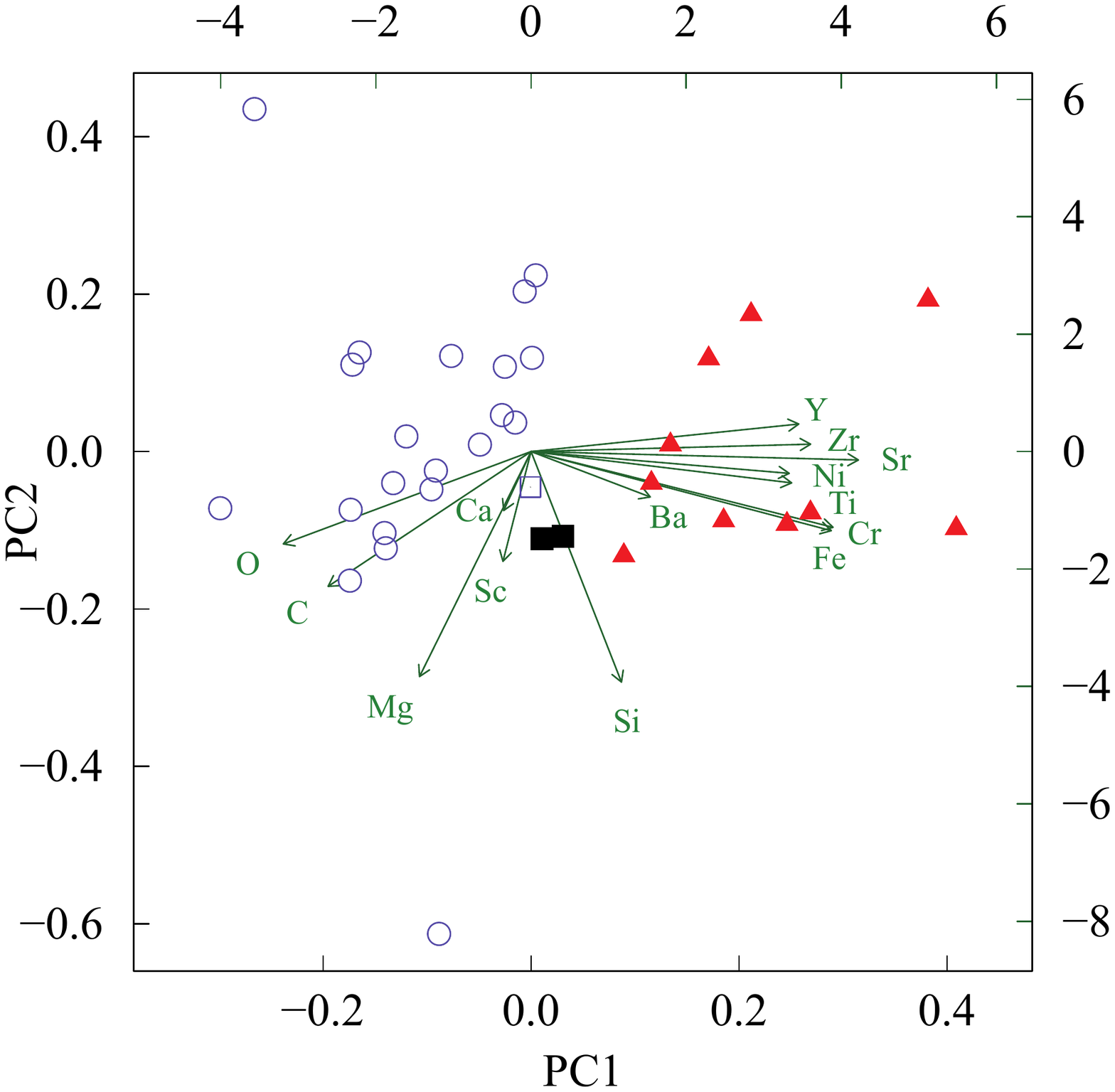}}
\caption{Observed sample and species projected on the first and second principal components (see text). The CP and normal stars are indicated by filled triangles and open circles respectively, according to the results from the cluster analysis. Filled squares represent the two uncertain objects and the open square stands for the probable normal star. The directions of the arrows show the relative loadings of the species on the first and second principal components.}
\label{PCA}
\end{figure}

The four criteria give rather consistent results. The distribution of $p_{14}$ shows slightly more separated and peaked groups than $p_{10}$, as a result of the additional discriminant species. 
A few objects show discrepant classifications: HD~219485 is classified as probably normal in Table\,\ref{table_classif} because the classification based on $p_{10}$ points it as a CP star whereas the other criteria indicate it is normal; two objects are classified as uncertain (HD~1439 and HD~219290) because the classification based on 10 species and the one based on 14 species produce contradictory results.

\onllongtab{
\begin{table}[!t]
\centering
\caption{Probabilities to be CP given by the four different criteria. Classification is deduced from the combination of four probabilities.}
\begin{tabular}{rrrrrc}
\hline\hline
HD~& $p_{10}$ & $f_{10}$ & $p_{14}$ & $f_{14}$ & Classification\\
\hline 
  1439 & 0.827 & 1 & 0.342 & 0 & uncertain        \\
 21050 & 0.107 & 0 & 0.012 & 0 & normal           \\
 25175 & 0.073 & 0 & 0.035 & 0 & normal           \\
 28780 & 0.287 & 0 & 0.348 & 0 & normal           \\
 30085 & 0.986 & 1 & 0.983 & 1 & CP               \\
 47863 & 0.004 & 0 & 0.001 & 0 & normal           \\
 58142 & 0.423 & 0 & 0.400 & 0 & normal           \\
 65900 & 0.999 & 1 & 0.994 & 1 & CP               \\
 67959 & 0.960 & 1 & 0.945 & 1 & CP               \\
 72660 & 0.999 & 1 & 1.000 & 1 & CP               \\
 73316 & 0.014 & 0 & 0.010 & 0 & normal           \\
 83373 & 0.854 & 1 & 0.640 & 1 & CP               \\
 85504 & 0.428 & 0 & 0.085 & 0 & normal           \\
 89774 & 0.166 & 0 & 0.051 & 0 & normal           \\
 95418 & 0.980 & 1 & 0.983 & 1 & CP               \\
104181 & 0.036 & 0 & 0.010 & 0 & normal           \\
107655 & 0.884 & 1 & 0.940 & 1 & CP               \\
127304 & 0.997 & 1 & 0.999 & 1 & CP               \\
132145 & 0.008 & 0 & 0.005 & 0 & normal           \\
133962 & 0.393 & 0 & 0.229 & 0 & normal           \\
145788 & 0.012 & 0 & 0.001 & 0 & normal           \\
154228 & 0.899 & 1 & 0.799 & 1 & CP               \\
156653 & 0.244 & 0 & 0.411 & 0 & normal           \\
158716 & 0.756 & 1 & 0.916 & 1 & CP               \\
172167 & 0.001 & 0 & 0.027 & 0 & normal           \\
174567 & 0.076 & 0 & 0.042 & 0 & normal           \\
176984 & 0.087 & 0 & 0.054 & 0 & normal           \\
196724 & 0.323 & 0 & 0.312 & 0 & normal           \\
198552 & 0.111 & 0 & 0.199 & 0 & normal           \\
199095 & 0.261 & 0 & 0.189 & 0 & normal           \\
219290 & 0.849 & 1 & 0.349 & 0 & uncertain        \\
219485 & 0.637 & 0 & 0.402 & 0 & probably normal  \\
223386 & 0.014 & 0 & 0.028 & 0 & normal           \\
223855 & 0.018 & 0 & 0.001 & 0 & normal           \\
\hline
\end{tabular}
\label{table_classif}
\end{table}
}

In order to check how separated are the two groups in the chemical abundance space, we perform a principal component analysis to display the data in fewer dimensions. Principal component analysis is a statistical method that expresses a set of variables, possibly correlated, as a set of linearly uncorrelated variables called principal components. The principal components are linear combinations of the original variables and are defined in such a way that they have the largest possible variance. We apply this method to our data and derive the first two principal components from the 14-species abundances. The determined elemental abundances are projected onto these two new axes in Fig.\,\ref{PCA}, as well as the 14 axes corresponding to the different species.
The first principal component (PC1) derived from the data explains 38.8\% of the total variation and the second one (PC2) explains 15.4\% of the variance in the 14-species abundances. Figure\,\ref{PCA} shows that species from titanium to zirconium have strong positive loadings on PC1, whereas oxygen and carbon have strong negative loadings. On the other hand, PC2 has strong loadings from magnesium and silicium. The original data for the 34 stars are expressed in the first two components and overplotted in Fig.\,\ref{PCA}. They are represented according to the classification resulting from the cluster analysis. The dichotomy is very clear along the first principal component, and the objects identified as ``uncertain'' in Table\,\ref{table_classif} lie between the two groups.

 Back in 2007, RZG used the release of the General Catalog of Ap and Am stars from \citet{Ren_91} to identify the chemically peculiar stars. This release is now outdated by \citet{RenMad09}. The 47 targets of our sample are not present in the previous version of the catalog, and the classification results can be compared with the content of the new release.
 Among these objects, nine are present in the catalog from \citet{RenMad09}, and are listed in Table\,\ref{comp_renson}. For two thirds of the common stars, both classifications are in agreement. 
\begin{table}[!t]
\centering
\caption{Comparison of the classification with the General Catalog of Ap and Am stars \citep{RenMad09}.}
\begin{tabular}{rcll}
\hline
\hline
HD~& \multicolumn{2}{c}{\citet{RenMad09}}& This work \\
\hline
 58142 &   & A0-A2    & normal \\
 72660 &   & A1-      & CP  \\
 83373 & ? & A1 Si    & CP \\
 85504 &   & A1 Mn    & normal \\
 95418 &   & A0- Ba Y & CP \\
107655 & ? & A1-      & CP \\
127304 & ? & B9 Si    & CP \\
145788 & ? & A1 Si    & normal \\
154228 & ? & A1 Si    & CP \\
\hline
\end{tabular}
\tablefoot{In the classification from \citet{RenMad09}, a dash indicates an Am star and no dash indicates an Ap star. The question mark indicates doubtful cases.}
\label{comp_renson}
\end{table}

On the other hand, four stars are classified as peculiar but are absent from the catalog built by \citet{RenMad09}: HD~30085, HD~65900, HD~67959, HD~158716. 
More details can be found in Appendix\,\ref{comments}.

\section{Discussion}  
  
  \label{sect_discussion}

\subsection{Abundance patterns}
  \label{subsect_patterns}

\begin{figure*}[!htp]
\centering
\resizebox{\hsize}{!}{\includegraphics[width=\textwidth,viewport=70 10 560 800,angle=-90]{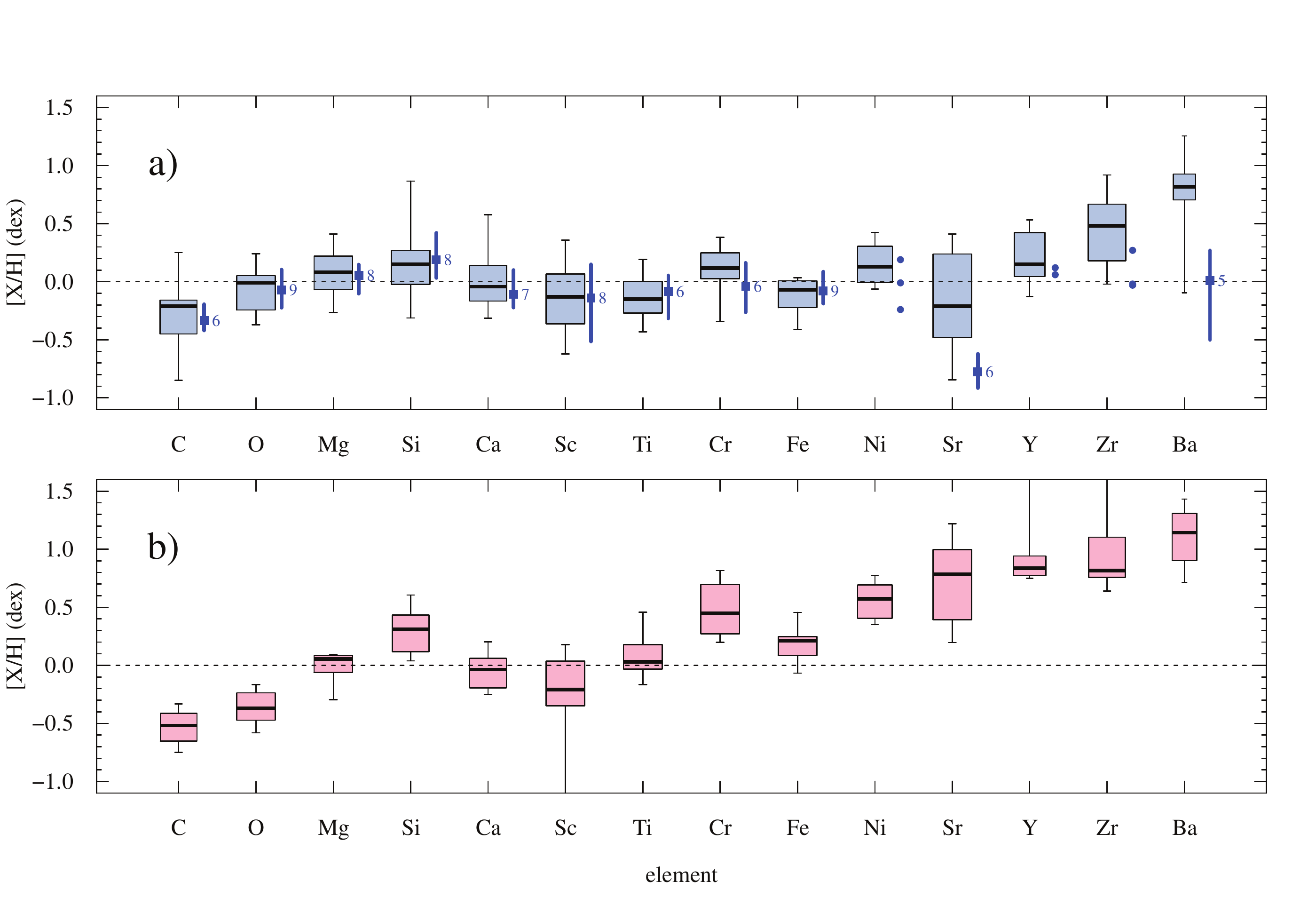}}
\caption{Abundance patterns are displayed as box and whisker plots for \textbf{(a)} the normal stars and \textbf{(b)} the CP stars. Boxes correspond to the 16th and 84th percentiles of the abundance distribution  for each element, thick horizontal lines represent the median value, and the whiskers span from the minimum to the maximum values. The width of the boxes is proportional to the square root of the number of values. The horizontal dashed line stands for the solar values. In the top panel, median values and 16th and 84th percentiles derived from the comparison stars are plotted on the right side of the boxes, as filled squares and thick vertical lines. The number of abundance determinations is indicated when five or more abundances are available, and all individual determinations are displayed when the number is less than five (see Sect.\,\ref{subsect_patterns}).}
\label{Pattern}
\end{figure*}

Figure\,\ref{Pattern} displays the median abundance patterns for the normal and peculiar stars resulting from our classification. Only fully agreeing classifications (all four criteria) are used to derive the median abundances, i.e. 21 normal stars and 10 CP stars. The dispersions are represented by the 16th and 84th percentiles, which in the Gaussian case correspond to the $\pm1$-$\sigma$ standard deviation.

The abundance pattern for the normal stars is compared with data from the Pleiades \citep{GenMor08} and from the Ursa Major moving group \citep{Mor05}. We select nine normal A-type stars corresponding to our observed effective temperature range ($\ensuremath{T_\mathrm{eff}}>8900$\,K): six stars are members of the Pleiades (HD~23763, HD~23948, HD~23629, HD~23632, HD~23489, HD~23387)
and three belong to Ursa Major (HD~1404, HD~12471, HD~209515). They are analyzed by the latter authors using the same method and the code from \citet{Taa95}, ensuring a comparison with homogeneous data.
Their median abundance pattern and the corresponding percentiles are overplotted in Fig.\,\ref{Pattern}a. It should be noted that this comparison sample is small and does not fully cover all the species. The number of available determinations for each species in the comparison sample is indicated in the plot.
The agreement between both patterns is very good but we notice discrepancies for strontium and the heavier elements.
The comparison stars have on average a larger rotational broadening: seven of them have $100\le\ensuremath{v\sin i}\le200$\,\ensuremath{\mbox{km}\,\mbox{s}^{-1}}.

As far as the abundance pattern for the CP stars is concerned, no object shows a significant underabundance in calcium and the Ca abundance is very similar between CP and normal stars. This tendency is possibly due to the fact that previously known CP stars are not part of our sample. We find that carbon and oxygen are underabundant, in agreement with \citet{RoyLat90}. The iron-peak elements and heavy elements are overabundant, compared to the normal stars.

By construction, the \ensuremath{v\sin i}\ range is limited in our sample. No clear trend of abundances versus \ensuremath{v\sin i}\ is detected in our data, neither on the whole dataset nor considering both groups separately.
  
\subsection{Rotational velocity distribution}
  Sections \ref{sect_rv}, \ref{sect_params} and \ref{sect_classif} show that our sample is still contaminated by CP stars and binary stars. These new identifications allow the analysis of the rotational velocity distribution of normal stars with a much cleaner sample. It can be noticed that in our sample, no CP stars were found with $\ensuremath{v\sin i}\gtrsim 45$\,\ensuremath{\mbox{km}\,\mbox{s}^{-1}}. The contamination by binary stars remains larger than by CP stars.

\begin{figure}[!t]
\centering
\resizebox{\hsize}{!}{\includegraphics[width=\textwidth]{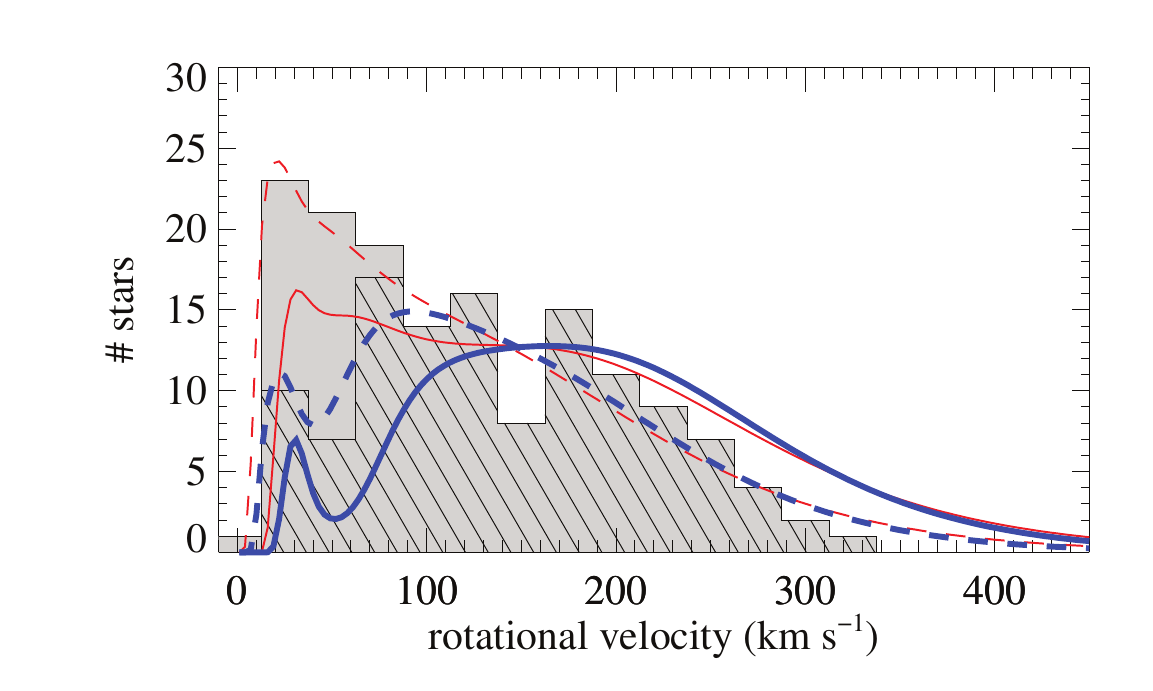}}
\caption{Distributions of rotational velocities: histograms are the observed \ensuremath{v\sin i}\ (hatched histogram corresponds to the cleaned sample); dashed lines stand for the smoothed distribution of projected rotational velocities; solid lines are the distributions of equatorial velocities. Thick lines correspond to the newly cleaned sample of 121 stars whereas thin lines represent the full sample of 151 stars.}
\label{distribs}
\end{figure}

 As defined in Sect.\,\ref{completeness}, our sample of 47 stars is the low \ensuremath{v\sin i}\ part of a larger sample of 151 normal A0--A1 stars representing an 80\%-complete, magnitude-limited volume. The distribution of rotational velocities of normal A0--A1 can be analyzed with this larger sample. The gray histogram in Fig.\,\ref{distribs} represents the distribution of their observed \ensuremath{v\sin i}.
 By removing the 30 stars identified as peculiar and/or binaries in the previous sections, a cleaned subsample of 121 stars is selected, and its distribution of \ensuremath{v\sin i}\ is showed by the hatched histogram (Fig.\,\ref{distribs}). 
The smoothed distributions of \ensuremath{v\sin i}\ is obtained by applying the method described in \citet{BonAzi97} and ported to R \citep{R}, and are shown by the dashed lines in Fig.\,\ref{distribs}. These distributions are then rectified from the projection effect to recover the distributions of equatorial velocities, assuming the rotation axes are randomly oriented (see RZG for details).

The high probability density for slow velocities disappears from the distribution in the cleaned sample, but a significant proportion (about 14\%) of the normal stars rotates slowly at $v\le 100$\,\ensuremath{\mbox{km}\,\mbox{s}^{-1}}. These results considerably reduce the presence of slowly rotating normal A0--A1 stars.
The overdensity at $v\le 50\,\ensuremath{\mbox{km}\,\mbox{s}^{-1}}$ represents 4\% of the distribution, i.e. five stars.

In the sample, the normal star with the lowest \ensuremath{v\sin i}\ is HD~145788 ($\ensuremath{v\sin i}=9.8$\,\ensuremath{\mbox{km}\,\mbox{s}^{-1}}\ as measured from the \'ELODIE spectrum). In the rectified distribution, there is no star rotating more slowly than $v=20$\,\ensuremath{\mbox{km}\,\mbox{s}^{-1}}. 

\section{Summary and conclusion}  
  
  \label{sect_conclusion}

\begin{figure}[!t]
\centering
\resizebox{\hsize}{!}{\includegraphics[viewport=100 130 470 700,angle=-90]{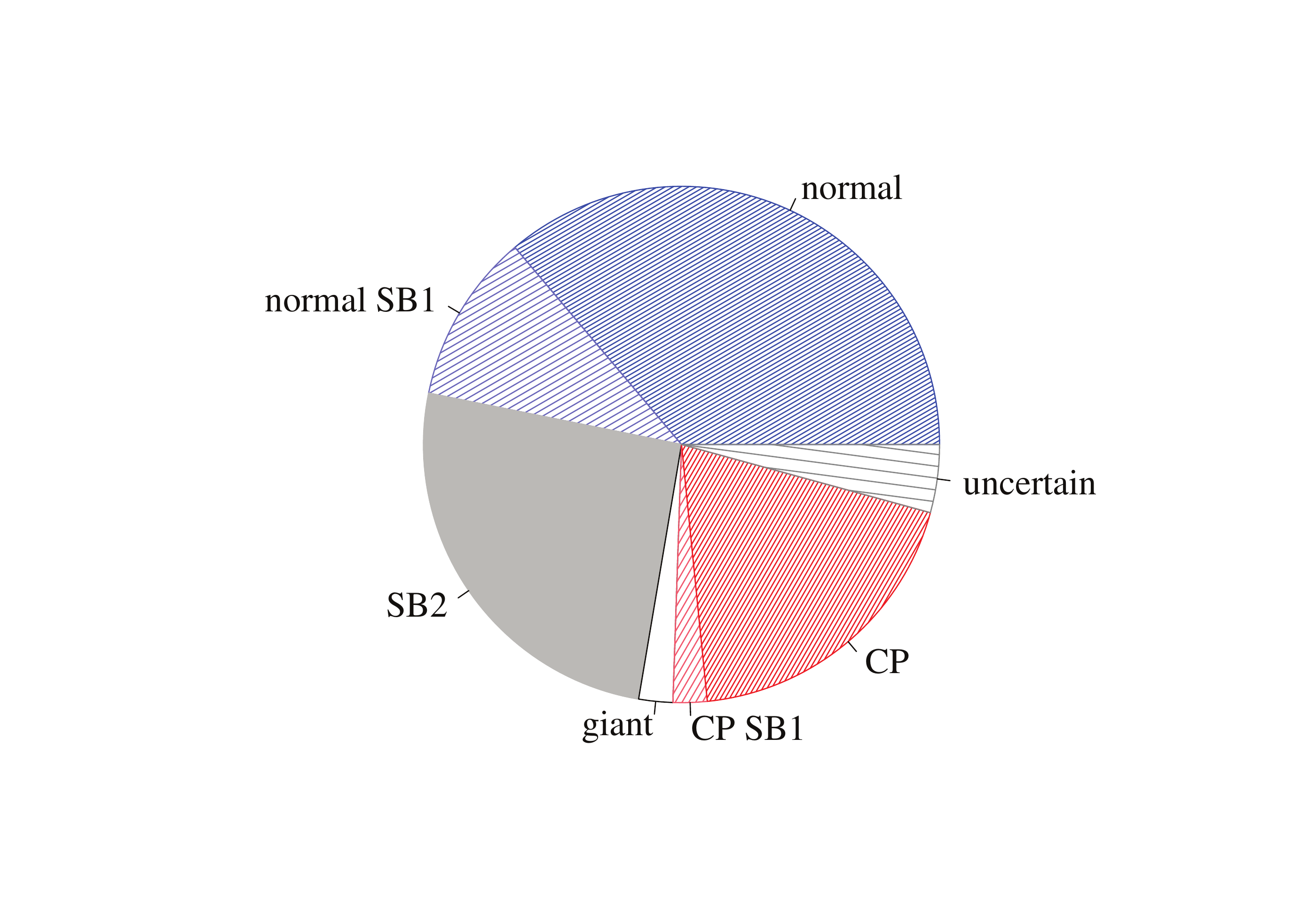}}
\caption{Resulting content of the sample of 47 A0--A1 stars.}
\label{piechart}
\end{figure}

This work provides the spectroscopic study of a sample of 47 A0--A1 stars, initially selected from \citet{Ror_02b,Ror_07} to be main-sequence, low \ensuremath{v\sin i}, normal stars. The analysis of the cross-correlation profiles, and the variation of radial velocities allow the identification of suspected spectroscopic binaries. The spectral synthesis and the determination of chemical abundances are used to identify chemically peculiar stars using a hierarchical classification. The results reveal that two thirds of the sample is composed of spectroscopic binaries and chemically peculiar stars, and only 17 stars turn out to be normal, showing no sign of multiplicity nor peculiarity. The final composition of the sample is given by the pie chart in Fig.\,\ref{piechart}.

  In the framework of a nearly complete, magnitude limited sample of 121 A0--A1-type normal stars, this implies that the distribution of equatorial velocities, under the assumption of randomly oriented rotation axes, is no longer dominated by a large fraction of slowly rotating stars. Only 14\% of the A0--A1 normal stars are rotating at $v\le 100$\,\ensuremath{\mbox{km}\,\mbox{s}^{-1}}, which correspond to about 16 objects, and 21\% are rotating at $v\le 120$\,\ensuremath{\mbox{km}\,\mbox{s}^{-1}}. This is to be compared with the corresponding proportions in the non-cleaned sample: 31\% and 37\% respectively (Fig.\,\ref{distribs}). The distribution is not bimodal as previously claimed by RZG, but the overlap with the distribution of equatorial velocity of CP stars seems real, contrary to the conclusion of \citet{AbtMol95} and \citet{Abt00}.

\begin{table}[!t]
\caption{List of HD~identifiers for the single normal stars. The ones written in boldface are suspected to be doubtful candidates from the comparison with literature data (see Appendix\,\ref{comments}).}
\label{normal}
\setlength{\tabcolsep}{3pt}
\centering
\begin{tabular}{rrrrrrr}
\hline\hline
 21050 &  25175 &  \textbf{28780} &  47863 &  \textbf{58142} &  73316 &  85504 \\
 89774 & 104181 & 132145 & 133962 & 145788 & 172167 & 198552 \\
\textbf{219485} & 223386 & 223855\\
\hline
\end{tabular}
\end{table}
  The 17 normal stars, spectroscopically identified in this paper, are listed in Table\,\ref{normal}. The objects were not extensively monitored in terms of radial velocity and spectroscopic binarity may still remain undetected in this sample. Possibly doubtful candidates are indicated in Table\,\ref{normal}, which could be either spectroscopic binaries or marginal CP stars (see Appendix\,\ref{comments}). The 17 normal stars will be more deeply investigated in a forthcoming paper to check whether signatures of gravity darkening due to fast rotation seen pole-on are present in the spectra. If rotation axes are randomly oriented, the probability to observe a star with $i\le 10\degr$ is about 1.5\%, which would produce just two stars among the 121 normal A0--A1 stars.
  
\begin{acknowledgements} 
We are very thankful to the referee for his/her appropriate and constructive suggestions and his/her proposed corrections to improve the paper.

This work has made use of BaSTI web tools (ver. 5.0.1). This research has made use of the VizieR catalog access tool, CDS, Strasbourg, France.

We are grateful to S.~Ilovaisky for providing calibration data corresponding to SOPHIE observations of Vega. 

\end{acknowledgements}


\Online


\begin{appendix}
\section{Comments on individual stars}
\label{comments}

\begin{figure*}[!ht]
\centering
\resizebox{\hsize}{!}{\includegraphics{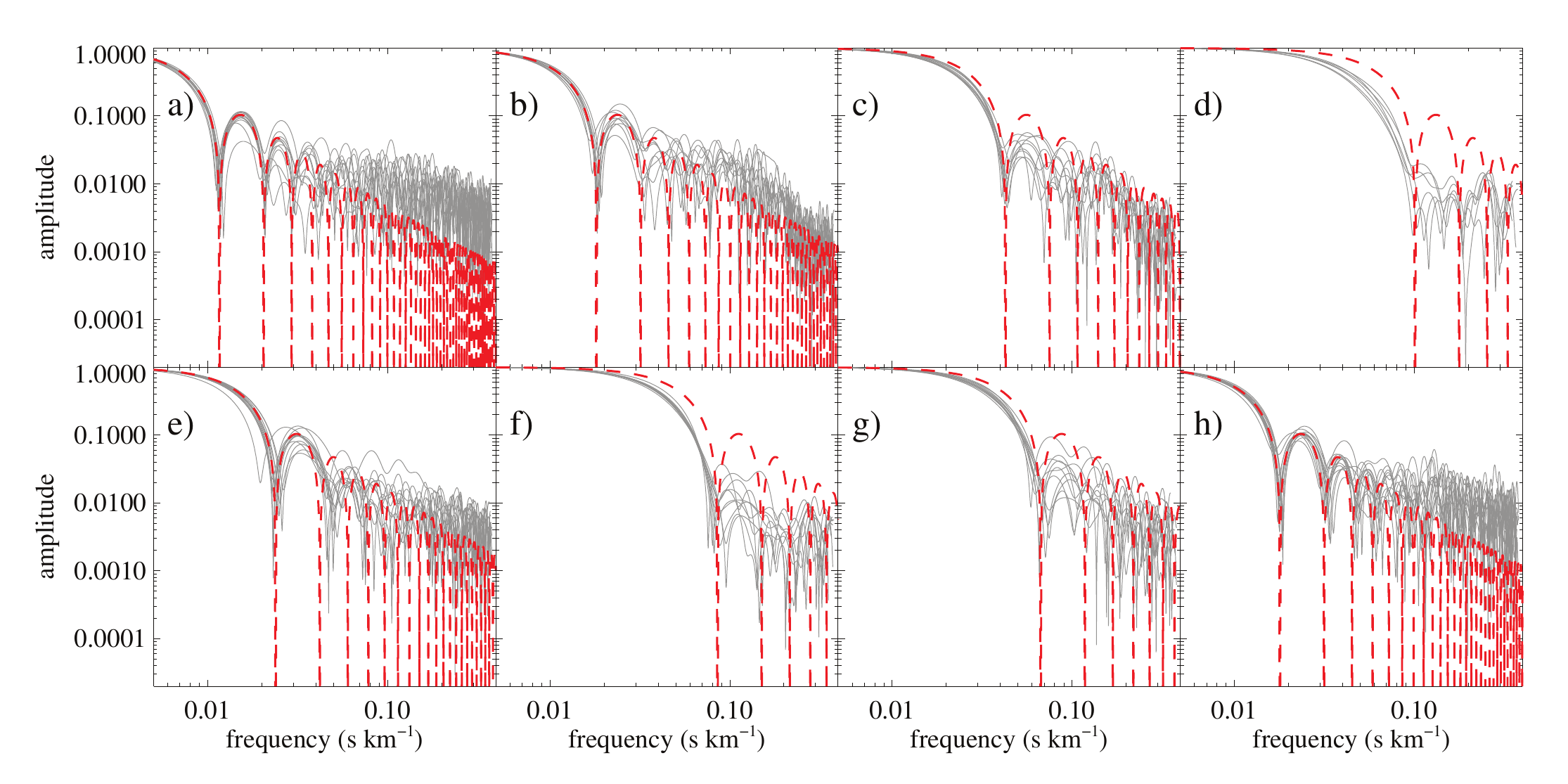}}
\caption{Individual line FT (solid lines) and theoretical rotation profile corresponding to the average \ensuremath{v\sin i}\ (thick dashed line) for the stars flagged as `0' by \citet{Raa_89}: 
\textbf{(a)} HD~25175, 
\textbf{(b)} HD~65900, 
\textbf{(c)} HD~67959, 
\textbf{(d)} HD~72660, 
\textbf{(e)} HD~85504, 
\textbf{(f)} HD~127304, 
\textbf{(g)} HD~145788, 
\textbf{(h)} HD~223386.
}
\label{FFTs}
\end{figure*}

All our targets are part of the sample studied by \citet{Dwy74}, and 25 of them are analyzed by \citet{Raa_89}. In their paper, \citet{Raa_89} derive \ensuremath{v\sin i}\ from Fourier profile analysis and flag the stars according to the agreement between the observed profile and a theoretical rotation profile (`1' when agreeing, `0' when disagreeing, in their Table\,3). Stars labeled as `0' by \citeauthor{Raa_89} are checked out with our data, and the Fourier profiles are plotted in Fig.\,\ref{FFTs}.

\begin{figure}[!ht]
\centering
\resizebox{\hsize}{!}{\includegraphics{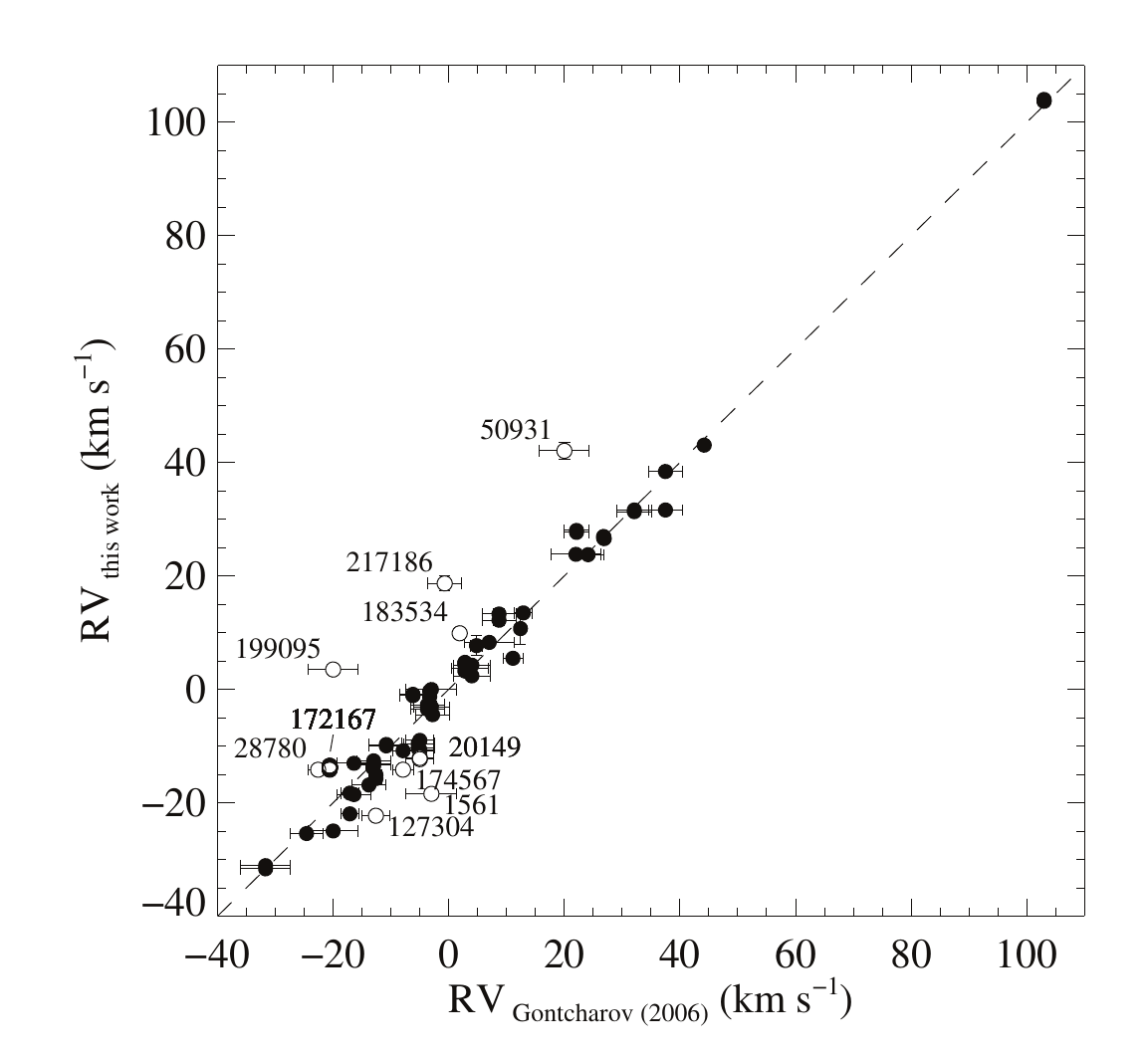}}
\caption{Comparison of radial velocities between individual values from Table\,\ref{indivspectra} and the compilation from \citet{Gov06}. The error bars are the internal errors and the dashed line is the one-to-one relation. The outliers are indicated by open symbols, together with their HD~number.
}
\label{VRcomp}
\end{figure}

\citet{Gov06} published a compilation of radial velocities and our individual measurements are compared with literature data. The 44 stars in common are plotted in Fig.\,\ref{VRcomp}. The Gaussian fit of the histogram of radial velocity differences gives a standard deviation $\sigma=2\,\ensuremath{\mbox{km}\,\mbox{s}^{-1}}$. Ten stars show differences larger than $3\,\sigma$. They are indicated in Fig.\,\ref{VRcomp}. The three stars of our sample (HD~40446, HD~119537 and HD~176984), that are not present in \citet{Gov06}, are already detected as binaries in our data.

\paragraph{\object{HD~1439}} could not be undoubtedly classified as a normal nor CP star, the derived memberships based on 10 and 14 elements giving contradictory results (Table\,\ref{table_classif}). It moreover lies in the ``uncertain'' zone in Fig.\,\ref{PCA}. The derived abundances in oxygen and magnesium are high enough to make it classified as a normal star using all 14 elements whereas the remaining pattern (Si, Ca, Sc, Cr, Fe, Sr, Y and Zr) is rather similar to a CP star. Its radial velocity observed 18 months apart does not show significant variation.

\paragraph{\object{HD~1561}} is variable in radial velocity, from our two observations as well as compared with \citet{Gov06} who gives a radial velocity of $-3\,\ensuremath{\mbox{km}\,\mbox{s}^{-1}}$ (Fig.\,\ref{VRcomp}).

\begin{figure}[!ht]
\centering
\resizebox{\hsize}{!}{\includegraphics{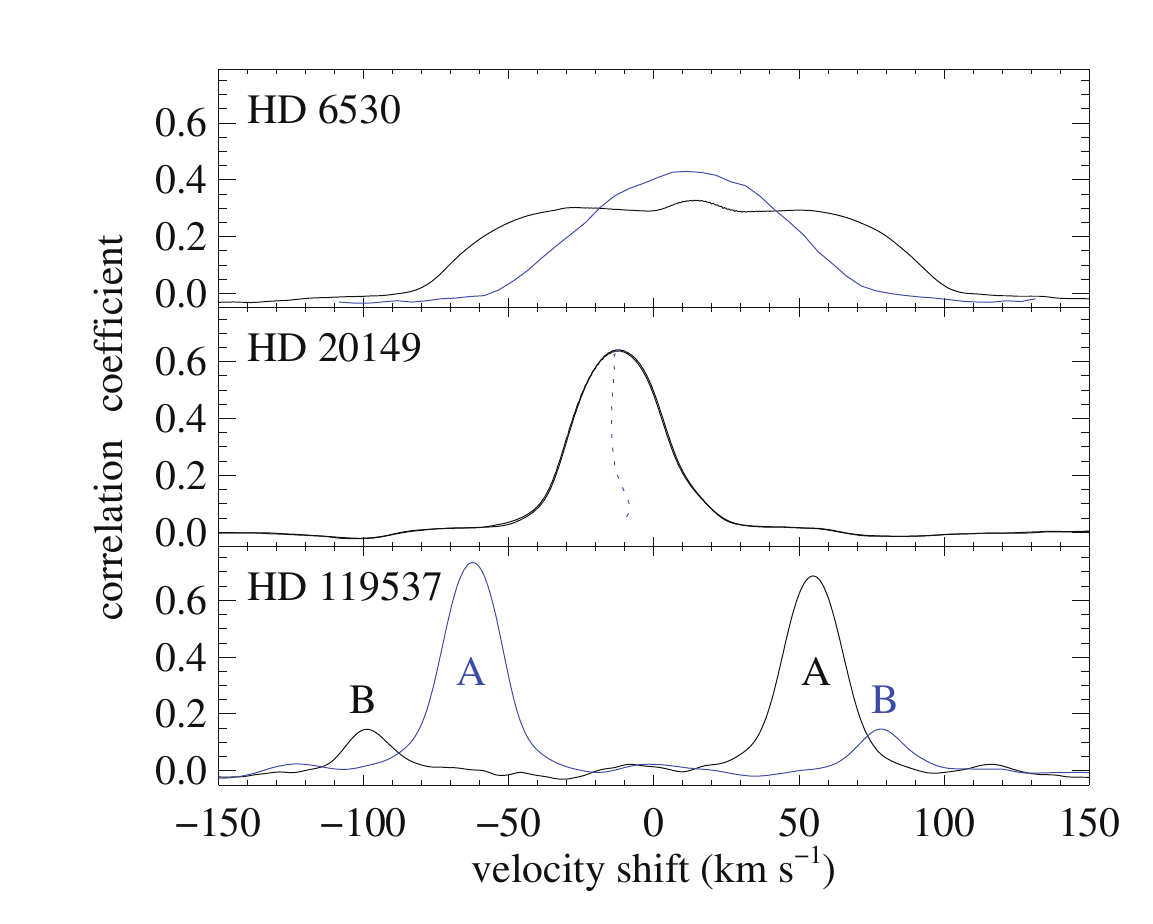}}
\caption{Cross-correlation functions for three suspected binary stars. \textbf{Top panel}: the CCF of the observed spectrum is overplotted to the one observed by \citet{Grr_99a}. \textbf{Middle panel}: the bisector of the CCF is displayed (dotted line) with a velocity scale enhanced by a factor 3, for the sake of clarity. \textbf{Bottom panel}: The two components in the composite CCF are labeled as `A' and `B'. 
}
\label{XCannex}
\end{figure}

\paragraph{\object{HD~6530}} is detected as a spectroscopic binary from its CCF (Fig.\,\ref{XC}). The large difference in the \ensuremath{v\sin i}\ measured using the spectral synthesis and the FT (Table\,\ref{tabstars}) is due to the composite spectrum. It has already been observed by \citet{Grr_99a} who derive the radial velocity and their spectrum is also used by \citet{Ror_02a} to derive the \ensuremath{v\sin i}. The CCF from this spectrum is overplotted in Fig.\ref{XCannex} (black solid line) to emphasize the binary nature of the star.

\paragraph{\object{HD~20149}} is flagged as a suspected binary due to the asymmetry of the CCF. The bisector is displayed in Fig.\ref{XCannex}. The star is labeled as variable in radial velocity in \citet{Gov06}, and the published value ($-5\,\ensuremath{\mbox{km}\,\mbox{s}^{-1}}$) is also significantly different from our determinations (Fig.\,\ref{VRcomp}).

\paragraph{\object{HD~21050}} is found to have a large projected rotational velocity by \citet{Dwy74} ($\ensuremath{v\sin i}=60\,\ensuremath{\mbox{km}\,\mbox{s}^{-1}}$) but all other determinations are very similar to our result \citep{Par_68,AbtMol95,Ror_02b}. Moreover, no spectral variation is detected in our high signal-to-noise observations, collected three years apart.

\paragraph{\object{HD~25175}} is suspected by \citet{Raa_89} to be a spectroscopic binary due to the large broadening and the disagreement the observed profile and a theoretical rotational profile. Their determination of \ensuremath{v\sin i}, in good agreement with ours, is much higher than the value derived by \citet{Dwy74} ($\le 40\,\ensuremath{\mbox{km}\,\mbox{s}^{-1}}$). Our Fourier profiles are plotted in Fig.\,\ref{FFTs}a and the agreement with the rotational profile is very good, suggesting that the broadening is dominated by rotation. Also both our spectra, observed one year apart, do not show any sign of radial velocity variation.
 
\paragraph{\object{HD~28780}} has a \ensuremath{v\sin i}\ significantly different from what \citet{Raa_89} find (41.3\,\ensuremath{\mbox{km}\,\mbox{s}^{-1}}). The radial velocity is moreover significantly different from the value published by \citet{Gov06}: $-22.6\,\ensuremath{\mbox{km}\,\mbox{s}^{-1}}$ (Fig.\,\ref{VRcomp}). Although no evidence of binarity is detected in our single observation, these differences suggest that this object could be a spectroscopic binary.

\paragraph{\object{HD~30085}} is discarded from \ensuremath{v\sin i}\ measurement by \citet{Raa_89} because of asymmetric line profiles. In our classification, it falls in the CP group and is newly detected as chemically peculiar.

\paragraph{\object{HD~33654}} is identified as a giant star from both its surface gravity and its luminosity (Table\,\ref{tabstars}), which disagrees with its luminosity class: A0V. It was previously classified as a B9III star \citep{Par_68}. It is indicated as an Ap Si star in \citet{RenMad09}.

\paragraph{\object{HD~39985}} is an outlier in our luminosity comparison (Fig.\,\ref{Barry}) and therefore suspected to be a binary star.

\paragraph{\object{HD~40446}} is flagged as a spectroscopic binary by \citet{Dwy74}. It is part of the sample studied by \citet{Ror_02b} who determined an uncertain \ensuremath{v\sin i}\ with a high external error ($\ensuremath{v\sin i}=27\mbox{:}\pm 5\,\ensuremath{\mbox{km}\,\mbox{s}^{-1}}$) due to a large dispersion in the values derived from individual lines.
Its binary nature is confirmed by the shape of its CCF in Fig.\,\ref{XC}.

\paragraph{\object{HD~46642}} is suspected to be a photometric variable star \citep{Kun_81}. Our radial velocity measurements reveal a variation hence this star is suspected to be a binary. 

\paragraph{\object{HD~47863}} shows a significant difference between our measurements of \ensuremath{v\sin i}, using spectral synthesis and Fourier profile. This object is however used as a reference star in speckle observations by \citet{Ari_97}, suggesting that it is a single star.

\paragraph{\object{HD~50931}} is detected as a spectroscopic binary from the distorted shape of its CCF. The radial velocity from \citet{Gov06} is moreover very different from our result: 20\,\ensuremath{\mbox{km}\,\mbox{s}^{-1}}\ (Fig.\,\ref{VRcomp}). The large difference in the \ensuremath{v\sin i}\ measured using the spectral synthesis and the FT (respectively 75 and 84\,\ensuremath{\mbox{km}\,\mbox{s}^{-1}}\ in Table\,\ref{tabstars}) is due to the composite spectrum.

\paragraph{\object{HD~58142}} is found to be a hot Am star by \citet{Adn94}, which disagrees with our classification. This object lies in the tail of the distribution of memberships (Table\,\ref{table_classif} and Fig.\,\ref{classif}) and could have been wrongly assigned to the ``normal'' group.

\paragraph{\object{HD~65900}} is suspected by \citet{Raa_89} to be a spectroscopic binary due to the large broadening and the disagreement between the observed profile and a theoretical rotational profile. This disagreement is not seen in our data (Fig.\,\ref{FFTs}b). In our classification, it falls in the CP group and is newly detected as chemically peculiar.

\paragraph{\object{HD~67959}} is found to disagree with a rotation profile by \citet{Raa_89}, but this is very probably due to its low \ensuremath{v\sin i}\ and the fact that instrumental broadening is not negligible. In our data (Fig.\,\ref{FFTs}c), the main lobe of the FT in observed profiles is well fitted by the theoretical rotation profile. In our classification, it falls in the CP group and is newly detected as chemically peculiar.

\paragraph{\object{HD~72660}} is a hot Am star \citep{Vae99}, which is consistent with our classification. Its low \ensuremath{v\sin i}\ makes the Fourier profiles dominated by the instrumental profile, both in \citet{Raa_89} and in Fig.\,\ref{FFTs}d. It is listed in Table\,\ref{binaries_deltavr} as variable in radial velocity, but the variation remains very small. \citet{Lat98} detects an asymmetry in the spectral lines that is attributed to a depth-dependent velocity field.

\paragraph{\object{HD~85504}} is flagged by \citet{Raa_89} as showing a profile disagreeing with a rotational broadening. In our data however (Fig.\,\ref{FFTs}e), the agreement with the theoretical rotational profile is good. This object is flagged in \citet{RenMad09} as an Ap\,Mn star, and manganese is not part of our analyzed chemical species. In \citet{AdnPio97}, it only appears slightly metal rich compared to other superficially normal stars with similar effective temperature.
This object is moreover known as high spacial velocity \citep{Mat70,AlndBr00} which may be a runaway star. It is a suspected variable star \citep{Kun_81}.

\paragraph{\object{HD~95418}} is an Am star \citep{Adn_11,Hil95} and our classification agrees with these results.

\paragraph{\object{HD~101369}} is suspected to be a spectroscopic binary from the shape of its CCF (Fig.\,\ref{XC}).

\paragraph{\object{HD~107655}} belongs to the open cluster Coma Ber. \citet{Gen_08} derived abundances, using the same method, and their values are in good agreement with our determinations, with larger differences for Sc and Sr. The dispersion in the Sc and Sr abundances derived by \citet{Gen_08} is high; they include more lines than in this study, especially lines located in the wings of Balmer lines. When restricting the comparison to lines in common, the agreement is much better (Table\,\ref{107655}).

\begin{table}[!ht]
\centering
\caption{Comparison of Sc and Sr abundances of \object{HD~107655} with \citet{Gen_08}. The mean abundances, relative to the Sun, are given, as well as the absolute abundances for each spectral line. Ellipsis dots (...) indicate when a spectral line has not been used.}
\label{107655}
\begin{tabular}{lcc}
\hline\hline
Element / line & this study & \citet{Gen_08} \\
\hline
[Sc/H] &  $0.178\pm0.066$ & $-0.2\pm0.11$ \\
4314.083\,\AA & ...            & 2.55 \\
4320.732\,\AA & ...            & 3.02 \\
4324.996\,\AA & ...            & 2.81 \\
4374.457\,\AA & 3.374 & 3.14 \\
5657.896\,\AA & 3.243 & 3.13 \\
\hline
[Sr/H] &  $0.197\pm0.150$ & $-0.23\pm0.20$ \\
4077.709\,\AA & ...            & 2.29 \\
4215.520\,\AA & 3.127 & 3.08 \\
\hline
\end{tabular}
\end{table}

\paragraph{\object{HD~119537}} is detected as a spectroscopic binary by \citet{Dwy74}. The SB2 nature is clearly visible in our CCFs and they are plotted for both observations in Fig.\ref{XCannex} together with the labels of the components (`A' being the component with the highest correlation peak). The radial velocities given in Table\,\ref{indivspectra} are the one corresponding to component A and the difference in radial velocity (A$-$B) is $153.4\,\ensuremath{\mbox{km}\,\mbox{s}^{-1}}$ in the first observation (2006-06-02) and $-140.9\,\ensuremath{\mbox{km}\,\mbox{s}^{-1}}$ on the second (2012-02-14).

\paragraph{\object{HD~127304}} is suspected by \citet{Raa_89} to be a SB2. Our derived radial velocity is moreover significantly different from the value published by \citet{Gov06}, $-12.6\,\ensuremath{\mbox{km}\,\mbox{s}^{-1}}$ (Fig.\,\ref{VRcomp}).

\paragraph{\object{HD~132145}} is suspected by \citet{Raa_89} to be a spectroscopic binary. The observed variation in radial velocity in our data is not significant taking the instrumental offset between both spectrographs into account.

\paragraph{\object{HD~133962}} is suspected to be a photometric variable star \citep{Kun_81}.

\paragraph{\object{HD~145647}} is suspected to be a binary star from the shape of the CCF (Fig.\,\ref{XC}).

\paragraph{\object{HD~145788}} is studied by \citet{Foi_09} who do not detect clear signatures of chemical peculiarity and believe this object is a normal star whose abundance pattern reflects the composition of its progenitor cloud. It is flagged by \citet{Raa_89} as showing a profile disagreeing with a rotational broadening, probably due to the small rotational broadening. In Fig.\,\ref{FFTs}g, the shape of the main lobes of the observed profiles agree with the theoretical rotation profile. This star was only observed with \'ELODIE and in our classification, it falls in the normal group. In the sample, it is the normal star with the smallest \ensuremath{v\sin i}.

\paragraph{\object{HD~156653}} is indicated as a spectroscopic binary by \citet{Dwy74}. Our radial velocity measurements show a ratio of the external error over the internal error just above the threshold used to identify variations in Table\,\ref{binaries_deltavr}. 

\paragraph{\object{HD~158716}} is suspected by \citet{Raa_89} to be a spectroscopic binary. In our classification, it falls in the CP group and is newly detected as chemically peculiar.

\paragraph{\object{HD~172167}} is, strangely enough, an outlier in Fig.\,\ref{VRcomp}, where \citet{Gov06} gives a radial velocity of $-20.6\,\ensuremath{\mbox{km}\,\mbox{s}^{-1}}$. Vega is not known to have a variable radial velocity on such a scale.

\paragraph{\object{HD~174567}} is suspected by \citet{Raa_89} to be a variable CP star (strong Si and variable Sr). It is classified as a normal star according to our measurements but we detect a variation in radial velocity, strengthened by the comparison with \citet{Gov06} in Fig.\,\ref{VRcomp}.

\paragraph{\object{HD~176984}} is suspected to be a spectroscopic binary from the variability of its radial velocity (Table\,\ref{binaries_deltavr}).

\paragraph{\object{HD~183534}} is detected as a spectroscopic binary from the shape of its CCF (Fig.\,\ref{XC}) and the difference in radial velocity with \citet{Gov06} in Fig.\,\ref{VRcomp}.

\paragraph{\object{HD~196724}} is suspected to be a spectroscopic binary from the variability of its radial velocity (Table\,\ref{binaries_deltavr}).

\paragraph{\object{HD~199095}} is detected as a spectroscopic binary by \citet{Dwy74}. Our measurements show a variation in radial velocity, confirmed when comparing with data from \citet{Gov06} in Fig.\,\ref{VRcomp}.

\paragraph{\object{HD~217186}} is considered to be a possible magnetic field candidate by \citet{Scr_08} after being associated with a ROSAT X-ray source \citep{ScrSct07} as a bona fide single star. This object is however suspected to be a binary from the shape of its CCF (Fig.\,\ref{XC}). We have only one spectrum for this object, but the comparison with \citet{Gov06}, in Fig.\,\ref{VRcomp}, suggests a variation in radial velocity.

\paragraph{\object{HD~219290}} is listed as ``uncertain'' in Table\,\ref{table_classif} and lies in the ``uncertain'' zone in Fig.\,\ref{PCA}. The high abundances in carbon and oxygen produce contradictory classifications when 10 and 14 are used. The remaining pattern (Si, Ca, Sc, Cr, Fe, Ni, Sr and Zr) is rather similar to a CP star. 

\paragraph{\object{HD~219485}} is listed as ``probably normal'' in Table\,\ref{table_classif}. It is marginally classified as a CP star based on the 10  ``classical'' elements. The abundances in Ni and Sr are however significantly higher than the median values in the normal star pattern displayed in Fig.\,\ref{Pattern}.

\paragraph{\object{HD~223386}} is flagged as disagreeing from a rotation profile by \citet{Raa_89}, but the agreement is very good in our data (Fig.\,\ref{FFTs}h).

\paragraph{\object{HD~223855}} is indicated as a spectroscopic binary by \citet{Dwy74}. It shows a significant difference between the measurements of \ensuremath{v\sin i}\ using spectral synthesis and Fourier profiles, and both values are significantly higher than the one derived by \citet{Dwy74}: $45\,\ensuremath{\mbox{km}\,\mbox{s}^{-1}}$.

\section{Linelist}

 The analyzed spectral lines are listed in Table\,\ref{linelist}, sorted by chemical element and central wavelength, together with the adopted oscillator strength.

\begin{table*}[!ht]
\caption{Atomic data and their reference for the lines used in our abundances determination. 
}
\centering
\begin{tabular}{lcrclcrclcrc}
\hline\hline
Element  & $\lambda$ (\AA) & $\log gf$ &Ref. &Element &  $\lambda$ (\AA)  & $\log gf$ &Ref. &Element &  $\lambda$ (\AA) & $\log gf$& Ref.\\
 \cline{1-12}
 \ion{C}{i} & 4932.049 & $-1.658$    & 1& \ion{Sc}{ii} & 4246.822 & $0.242$   & 1& \ion{Ni}{i}  & 4470.472 & $-0.310$ & 3 \\
\ion{C}{i} & 5052.167 & $-1.303$    & 1& \ion{Sc}{ii} & 4374.457 & $-0.418$  & 1& \ion{Ni}{i}  & 4480.561 & $-1.491$ & 3 \\
\ion{C}{i} & 5380.337 & $-1.616$    & 1& \ion{Sc}{ii} & 4670.407 & $-0.576$  & 1& \ion{Ni}{i}  & 4490.049 & $-2.108$ & 3 \\
\ion{C}{i} & 5793.120 & $-2.063$    & 1& \ion{Sc}{ii} & 5031.021 & $-0.400$  & 1& \ion{Ni}{i}  & 4490.525 & $-2.324$ & 3 \\
\ion{C}{i} & 5800.602 & $-2.337$    & 1& \ion{Sc}{ii} & 5239.813 & $-0.765$  & 1& \ion{Ni}{i}  & 4604.982 & $-0.250$ & 3 \\
\cline{1-4}
\ion{O}{i} & 5330.726 & $-2.416$    & 1& \ion{Sc}{ii} & 5526.790 & $0.020$   & 1& \ion{Ni}{i}  & 4648.646 & $-0.100$ & 3 \\
\ion{O}{i} & 5330.735 & $-1.570$    & 1& \ion{Sc}{ii} & 5657.896 & $-0.603$  & 1& \ion{Ni}{i}  & 5080.528 & $0.330$  & 3 \\
\cline{5-8}
\ion{O}{i} & 5330.741 & $-0.983$    & 1& \ion{Ti}{ii} & 4163.644 & $-0.130$  & 4& \ion{Ni}{i}  & 5099.927 & $-0.100$ & 3 \\
\ion{O}{i} & 6155.961 & $-1.363$    & 1& \ion{Ti}{ii} & 4287.873 & $-1.790$  & 4& \ion{Ni}{i}  & 5476.904 & $-0.890$ & 1 \\
\cline{9-12}
\ion{O}{i} & 6155.961 & $-1.363$    & 1& \ion{Ti}{ii} & 4417.714 & $-1.190$  & 4& \ion{Sr}{ii} & 4215.520 & $-0.169$ & 1 \\
\cline{9-12}
\ion{O}{i} & 6155.971 & $-1.011$    & 1& \ion{Ti}{ii} & 4443.801 & $-0.720$  & 4& \ion{Y}{ii}  & 4883.684 & $0.070$  & 3 \\
\ion{O}{i} & 6155.989 & $-1.120$    & 1& \ion{Ti}{ii} & 4468.492 & $-0.600$  & 5& \ion{Y}{ii}  & 4900.120 & $-0.090$ & 3 \\
\ion{O}{i} & 6156.737 & $-1.487$    & 1& \ion{Ti}{ii} & 4501.270 & $-0.770$  & 4& \ion{Y}{ii}  & 5087.416 & $-0.170$ & 3 \\
\cline{5-8}
\ion{O}{i} & 6156.755 & $-0.898$    & 1& \ion{Cr}{ii} & 4592.049 & $-1.217$  & 5& \ion{Y}{ii}  & 5200.406 & $-0.570$ & 3 \\
\cline{9-12}
\ion{O}{i} & 6156.778 & $-0.694$    & 1& \ion{Cr}{ii} & 4616.629 & $-1.291$  & 6& \ion{Zr}{ii} & 4156.240 & $-0.776$ & 3 \\
\ion{O}{i} & 6158.149 & $-1.841$    & 1& \ion{Cr}{ii} & 4634.070 & $-0.990$  & 3& \ion{Zr}{ii} & 4161.210 & $-0.720$ & 3 \\
\ion{O}{i} & 6158.172 & $-0.995$    & 1& \ion{Cr}{ii} & 4812.337 & $-1.995$  & 3& \ion{Zr}{ii} & 4208.980 & $-0.460$ & 3 \\
\ion{O}{i} & 6158.187 & $-0.409$    & 1& \ion{Cr}{ii} & 5237.329 & $-1.160$  & 5& \ion{Zr}{ii} & 4496.960 & $-0.810$ & 7 \\
\cline{1-4}
\cline{9-12}
\ion{Mg}{ii} & 4427.994 & $-1.201$  & 1& \ion{Cr}{ii} & 5308.440 & $-1.810$  & 5& \ion{Ba}{ii} & 4554.029 & $0.163$  & 3 \\
\ion{Mg}{ii} & 4481.126 & $0.730$   & 2& \ion{Cr}{ii} & 5313.590 & $-1.650$  & 5& \ion{Ba}{ii} & 4934.076 & $-0.156$ & 8 \\
\cline{5-8}
\ion{Mg}{ii} & 4481.150 & $-0.570$  & 2& \ion{Fe}{ii} & 4273.326 & $-3.258$  & 5& \ion{Ba}{ii} & 6141.713 & $-0.810$ & 1 \\
\ion{Mg}{ii} & 4481.325 & $0.575$   & 2& \ion{Fe}{ii} & 4296.572 & $-3.010$  & 5&                                     \\
\cline{1-4}
\ion{Si}{ii} & 5041.024 & $0.174$   & 1& \ion{Fe}{ii} & 4416.830 & $-2.600$  & 5&                                     \\
\ion{Si}{ii} & 5055.984 & $0.441$   & 1& \ion{Fe}{ii} & 4491.405 & $-2.690$  & 5&                                     \\
\ion{Si}{ii} & 5056.317 & $-0.535$  & 1& \ion{Fe}{ii} & 4508.288 & $-2.210$  & 5&                                     \\
\ion{Si}{ii} & 5466.432 & $-0.190$  & 3& \ion{Fe}{ii} & 4515.339 & $-2.490$  & 5&                                     \\
\ion{Si}{ii} & 5688.817 & $0.106$   & 1& \ion{Fe}{ii} & 4520.224 & $-2.600$  & 5&                                     \\
\ion{Si}{ii} & 5978.930 & $-0.061$  & 1& \ion{Fe}{ii} & 4522.634 & $-2.030$  & 5&                                     \\
\cline{1-4}
\ion{Ca}{ii} & 4489.179 & $-0.726$  & 3& \ion{Fe}{ii} & 4541.524 & $-3.050$  & 5&				      \\
\ion{Ca}{ii} & 4489.179 & $-2.157$  & 3& \ion{Fe}{ii} & 4555.890 & $-2.290$  & 5&				      \\
\ion{Ca}{ii} & 4489.179 & $-0.613$  & 3& \ion{Fe}{ii} & 4582.835 & $-3.100$  & 5&				      \\
\ion{Ca}{ii} & 5001.479 & $-0.517$  & 1& \ion{Fe}{ii} & 4656.981 & $-3.630$  & 5&				      \\
\ion{Ca}{ii} & 5019.971 & $-0.257$  & 1& \ion{Fe}{ii} & 4666.758 & $-3.330$  & 5&				      \\
\ion{Ca}{ii} & 5021.138 & $-1.217$  & 1& \ion{Fe}{ii} & 4923.927 & $-1.320$  & 5&				      \\
\ion{Ca}{ii} & 5285.266 & $-1.153$  & 1& \ion{Fe}{ii} & 5197.577 & $-2.100$  & 5&				      \\
\ion{Ca}{ii} & 5307.224 & $-0.853$  & 1& \ion{Fe}{ii} & 5276.002 & $-1.940$  & 5&				      \\
             &          &           &  & \ion{Fe}{ii} & 5316.615 & $-1.850$  & 5&				      \\
\hline
\end{tabular}
\label{linelist}
\tablebib{
(1) NIST; (2) \cite{BinLuk48}; (3) Kurucz (\url{http://kurucz.harvard.edu/LINELISTS/GFALL}); (4) \cite{Pig_02};
(5) \cite{Fur_88}; (6) \cite{SitLat90}; (7) \cite{Bit_81}; (8) \cite{MisWie69}.
}
\end{table*}
\end{appendix}

\end{document}